\documentclass[12pt]{amsart}
\usepackage{geometry} % see geometry.pdf on how to lay out the page. There's lots.
\usepackage{graphicx}

\usepackage{color,soul}
\setul{1ex}{0.8ex}
\definecolor{orange}{rgb}{1,0.5,0}
\setulcolor{orange}

\usepackage{ulem}

 \usepackage[small]{caption}
\usepackage{epstopdf}
\usepackage{subfigure}
\usepackage{float}
\newfloat{figure}{tbp}{lgr}   % optional numbering with [section] or [chapter]
\floatname{figure}{Supplementary Figure}
\newfloat{figure2}{tbp}{lgr}   % optional numbering with [section] or [chapter]
\floatname{figure2}{Figure}
\title{Standing genetic variation and the evolution of drug resistance in HIV}
\author{Pleuni S. Pennings}
\begin{document}

\maketitle
Harvard University, Department of Organismic and Evolutionary Biology,
4094 Biological Laboratories, 
16 Divinity Avenue, Cambridge,
MA 02138 USA 

\tableofcontents

\section{Abstract}
Drug resistance remains a major problem for the treatment of HIV. Resistance can occur due to mutations that were present before treatment starts or due to mutations that occur during treatment. The relative importance of these two sources is unknown. 
We study three different situations in which HIV drug resistance may evolve: starting triple-drug therapy, treatment with a single dose of nevirapine and interruption of treatment. For each of these three cases good data are available from literature, which allows us to estimate the probability that resistance evolves from standing genetic variation. Depending on the treatment we find probabilities of the evolution of drug resistance due to standing genetic variation between $0$ and $39\%$. For patients who start triple-drug combination therapy, we find that drug resistance evolves from standing genetic variation in approximately 6\% of the patients. We use a population-dynamic and population-genetic model to understand the observations and to estimate important evolutionary parameters. We find that both, the effective population size of the virus before treatment, and the fitness of the resistant mutant during treatment, are key-parameters that determine the probability that resistance evolves from standing genetic variation. Importantly, clinical data indicate that both of these parameters can be manipulated by the kind of treatment that is used. 

\section{Introduction}
For most HIV patients, treatment with modern antiretroviral therapy leads to a rapid decline of viral load (VL) of several orders of magnitude. However, when the virus acquires resistance to one or more drugs, treatment can fail.  
It is still an open question whether the mutations responsible for resistance originate usually from standing genetic variation (also referred to as pre-existing mutations or minority variants), or from new mutations which occur during therapy. In fact, there is no biological system for which the relative role of pre-existing and new mutations is well known (Barret 2008). The case of HIV drug resistance may serve as a good case-study for evolutionary biologists. In addition, understanding the role of pre-existing drug resistance mutations may help guide the design of better treatment strategies to reduce the risk that resistance evolves in patients. 

In this paper we will look at the establishment of drug resistance mutations in three different situations: (1) when triple-drug therapy (ART) is started for the first time, (2) when pregnant women are treated with a single dose of nevirapine to prevent infection of the baby during birth and (3) when standard therapy is interrupted and restarted. We will argue that standing genetic variation plays a crucial role in each of these cases. We find that the probability that resistance mutations become established in each of these cases can be understood by using a simple population genetic model.  

For readers who are not familiar with HIV, it is important to know that the genotype-phenotype map for drug resistance in HIV is very well known. Lists of the important resistance mutations for each drug are published (e.g. in the International AIDS SocietyÐUSA drug resistance mutations list, Johnson 2010), so that doctors can compare the genotype of the virus of a patient before treatment with this list to decide which drugs to prescribe. The aim of treatment is to achieve viral suppression. If treatment fails despite adherence to the regimen, a second genotypic test will be performed to see whether the virus has acquired new resistance mutations. Since the second half of the 1990s, treatment is usually with a combination of three drugs, which are chosen such that mutations which confer resistance against one of the drugs do not confer cross-resistance against the other two drugs. 
%The reasoning behind the use of three drugs is that it would make it harder for the virus to acquire resistance, because in order to escape from drug pressure the virus would need to acquire three resistance mutations at the same time. 
Soon it became clear that triple-drug therapy was an enormous success and saved the lives of many HIV patients (Mocroft 1998).
One reason why therapy with three drugs works better than treatment with one or two drugs is that the rate at which resistance evolves is slower when patients are treated with three drugs (Deeks 1998). It is commonly thought that resistance does not evolve in patients on triple-drug therapy because it would require a viral particle to acquire three mutations at the same time. 
However, in patients who are treated with triple-drug therapy, it is often observed that resistance against one of the drugs evolves, at least initially. Data from several cohort studies in different parts of the world, such as from Canada (Harrigan et al 2005) and the UK (UK CHIC cohort study, Cozzi-Lepri 2005) clearly show that in most patients who fail therapy due to resistance, the virus is resistant against one of the drugs and almost never against all three. The UK study, for example, reports that out of 4306 patients who started therapy between 1996 and 2003, after two years of therapy, 13\% have drug resistance, but less than half of these patients (6\%) have resistance against more than one class of drugs and a only small number of patients (1\%) have resistance against 3 classes of drugs, even though all patients of this cohort were treated with three classes of drugs. These data show that treatment can fail due to resistance against one of the drugs in a regimen. The viruses that have acquired resistance against two or three classes of drugs may have acquired several drug resistance mutations at the same time or they may have acquired resistance against one drug first and subsequently acquired resistance against the other classes. In order to distinguish between these two possibilities, a very dense sampling of a patient's viral population would be needed. For now, we will assume that in most patients the accumulation of resistance mutations is a gradual process.

For many common drugs, especially reverse transcriptase inhibitors, a single mutation can confer resistance against the drug and only a small number of mutations is responsible for resistance in most patients. For example, resistance against the drug nevirapine is almost always due to one of two amino acid changes, namely K103N or Y181C in the reverse-transcriptase gene (Paredes et al 2010). Because of the importance of a small number of mutations, several studies have investigated whether these mutations are present in untreated patients due to transmitted drug resistance or due to spontaneous mutation. Recent studies have used allele-specific PCR and related methods to determine the frequency of several important mutations in untreated patients. Low-frequency drug resistance mutations (DRMs), likely due to spontaneous mutation (and not transmitted from other patients) were detected in up to 40\% of patients (see Gianella and Richman 2010 for an overview). The detection of drug resistance mutations in untreated patients, together with the knowledge that a single mutation can confer resistance against a drug and allow viral escape, suggest that pre-existing resistance mutations (or standing genetic variation in the population genetic jargon) may play an important  role in the evolution of drug resistance in HIV. 

Throughout the paper, we will assume that a single mutation can allow viral escape and we focus on the probability that such a first drug resistance mutation becomes established (i.e., it reaches such a frequency that it can be expected to become the majority variant unless treatment is stopped or changed quickly). What happens after a first mutation has become established, or how fast such an established mutation wanes in the absence of treatment are important questions, but they fall outside the scope of this study. In this paper, `triple-drug therapy" refers to treatment with two drugs of the class NRTI plus either an NNRTI or an unboosted PI (for a list of abbreviations in the paper, see Table 2). The results are likely to be different for other drug combinations.

\emph{Starting therapy.} When a patient starts therapy for the first time, one would expect that there should be a substantial probability that drug resistance evolves due to pre-existing DRMs. Indeed, recent studies have shown that the presence of drug resistance mutations at low frequency ($<$1\%) increases the risk that treatment fails (e.g., Johnson 2008, Geretti 2009, Paredes 2010, see Li et al 2011 for a review). However, the situation is not as simple as one may hope: even if no pre-existing DRMs can be detected, resistance mutations may become established quickly, and even if DRMs are detected, treatment is still successful in the majority of patients. We will attempt to understand those observations using population genetic theory. 
Other authors have looked at the question of pre-existing DRMs previously (e.g., Bonhoeffer and Nowak 1998, Ribeiro and Bonhoeffer 2000), however, it is worth reconsidering the topic. First of all, we now have a wealth of data available for pre-existing DRMs and the establishment of drug resistance mutations in HIV patients, and secondly, we now have a better theoretical framework to consider the role of standing genetic variation for adaptation (Hermisson and Pennings, 2005).    

\emph{Prevention of mother to child transmission (PMTCT).} Pregnant women in low resource settings are often treated with a single dose of the non-nucleoside reverse transcriptase inhibitor neverapine when labor starts. Single dose nevirapine (sdNVP) is the cheapest and simplest way to reduce the probability of mother-to-child-transmission, but it is shown to lead to the establishment of drug resistance mutations in the mothers and the babies. In a meta-analysis, Arrive et al (2007) found that, in 7 different studies, on average 44\% of the patients treated with sdNVP had detectable NVP resistance mutation several weeks after the treatment. The presence of such mutations makes future treatment of these women harder (Lockman et al 2007). To avoid the establishment of resistance mutations, several alternative strategies are used in combination with sdNVP. We will use the same population genetic framework as in the other two cases to try to understand why sdNVP leads to establishment of resistance mutations in so many patients, and how this can be avoided. In the current study we will only focus on the probability that NVP resistance mutations become established during treatment for PMTCT. The issue of how these mutations wane and possibly resurface when treatment is started again is important and interesting but falls outside the scope of the current paper.

\emph{Treatment interruptions.} It was long suspected that treatment interruptions lead to drug resistance. Indeed, cohort studies show that treatment interruptions due to non-adherence are associated with faster accumulation of drug resistance mutations (Tam et al 2008, Gardner 2010, Lima 2010). Clear evidence that treatment interruptions of at least a couple of weeks lead to the establishment of resistance mutations comes from clinical trials (e.g., Yerly 2003, Danel 2006, 2009) which were done in a time when it was believed that treatment interruptions may be beneficial for patients. 
In 2006 the SMART trial was stopped because treatment interruptions were shown to have a negative effect on patients' health (El-Sadr 2008). However, treatment interruptions still occur, for example, when a patient is forgetful or is unable to purchase drugs due to financial or logistic barriers. It is important to understand how treatment interruptions lead to resistance and whether this effect can be avoided. 

The main idea that currently governs the thinking about treatment interruptions and resistance is that insufficient drug-levels allow for replication and, at the same time, select for resistance (e.g., Taylor 2007, Fox 2008, Gardner 2010). This effect is aggravated when drugs that are part of combination therapy have very different half-lifes, so that interrupting combination therapy can result in effective monotherapy. It is generally believed that this ``tail of monotherapy" is the main reason why treatment interruptions lead to drug resistance. However, several observations are not compatible with the ``tail" hypothesis. For example,  Fox et al (2008) found no significant difference in the number of resistance mutations after simultaneous, ``staggered" or ``switched" treatment interruptions in patients from the SMART trial (a ``staggered" stop means that the long half-life drug is interrupted several days before the other drugs and a ``switched" stop means that before interrupting, patients switch to a regimen with only short half-life drugs). 
In addition, the ``tail" hypothesis fails to explain why treatment interruptions increase the risk of resistance in patients on protease inhibitor-based (PI) regimens which do not have long half-lifes (Dybul 2003, Yerly 2003, Arnedo-Valero 2005, Henry 2006, Ruiz 2007, Darwich 2008). Another explanation is needed to understand the observed patterns. 

When treatment is interrupted, the viral load rapidly increases until it has reached its original level after approximately four weeks (Davey et al 1999). Basic population genetics tells us that such population growth also leads to an increase in the probability that DRMs are present. When treatment is started again, selection may work on such pre-existing mutations, which provides a simple explanation for how treatment interruptions lead to the establishment of resistance mutations. 

In this paper we will attempt to explain the observed patterns by considering selection on pre-existing variation and selection on new mutations. 
Throughout the paper, we use a mathematical model for adaptation from standing genetic variation which we developed previously (Hermisson and Pennings, 2005) and forward-in-time, individual-based computer simulations. The model captures mutation, drift and selection, including changing selection pressures (due to stopping and starting of therapy) which lead to changes in population size. Because we only focus on the establishment of the first drug resistance mutation, we can ignore epistatic interactions between different drug-resistance mutations and recombination. In each of the three cases of interest, we use published data on the percentage of patients with established drug resistance mutations to estimate important parameter values (for starting ART or sdNVP) and to predict outcomes (for treatment interruptions). 

%\pagebreak

\section{Model, assumptions and fixation probability of a drug resistance mutation}
The model we use in this paper describes the population dynamics and population genetics of a panmictic viral population in a single patient. Details of the model can be found in the supplementary material. We assume that as long as the patient is not on anti-retroviral therapy, the viral population will be stable at population size $N_u$ (u for untreated). Drugs reduce the fitness of the wildtype virus to below $1$ so that the population will shrink. We assume that there is a large reservoir of latently infected cells of which a fixed number ($I$) become activated per generation, so that the virus can not die out. Drug resistant virus can be created by mutation and is assumed to be resistant against one of the drugs in the treatment regimen. If the patient is not taking drugs, the drug resistant virus is less fit than the wildtype by a factor $C_{rel}$  (relative cost of the resistant virus), but if the patient is taking drugs, the resistant virus has a fitness that is higher than $1$ ($Fm_{ART}>1$), whereas the wildtype has a fitness lower than $1$ ($Fwt_{ART}<1$). In reality, there may also be resistance mutations that confer resistance against one of the drugs, but that do not lead to a fitness higher than $1$. Such mutations will quickly die out and can safely be ignored in the model. Throughout the paper we focus on the the processes that allow a first major drug resistance mutation to become established in the patient. Patients are assumed to be ART-naive and have no transmitted drug resistance. 

Evolutionary biologists have long known that most mutations will be lost by genetic drift even if they confer a fitness benefit (Haldane 1927). This is also true for drug resistance mutations (DRMs) in patients on anti-retroviral therapy, although it is all too often ignored in drug resistance studies. The clinical relevance of this old result has recently become very clear. It was found in several studies that even though low frequency DRMs increase the risk of treatment failure and establishment of drug resistance, the majority of patients with detected low frequency DRMs will respond well to treatment (Paredes et al 2010). This result shows that DRMs can die out, even if they have reached frequencies high enough to be detected. The reason is probably that most viral particles will not infect any new cells and produce no new viral particles, even if, on average, they produce more than $1$. Because of stochastic effects, not every particle will have the same number of offspring and if some have many, then others must have $0$. 
 
The probability that a DRM becomes established in the patient depends on the number of copies that are present, the average number of offspring of the drug resistant particles and the variance in offspring number. 
Traditionally, fixation or establishment probabilities are calculated using the relative fitness difference between the mutant and the wildtype, but in the case of HIV it is more useful to use the fitness of the mutant virus to calculate its establishment probability. 
The reason is that anti-retroviral therapy works so well that wildtype fitness may be very low (much lower than 1). In such case fitness of the mutant may not be related to the fitness of the wildtype and because the wildtype cannot grow, the two types do not compete for resources. In other words, the mutant can occupy a niche that is not occupied by the wildtype. In those cases the establishment probability of the mutant will be approximately 
$P_{est}\approx\frac{2*(Fm-1)}{\sigma}$
where $\sigma$ is the variance in offspring number. In the simulations and throughout this paper, we use the variance effective population size, in which case one can assume that $\sigma = 1$, so that 

\begin{equation}
\label{PfixSimple3}
P_{est}\approx2\;(Fm-1)
\end{equation}

It is important to realize that if the establishment probability of a DRM depends on its absolute fitness, anything that reduces its fitness will reduce the establishment probability. For example, if a drug that is added to a regime reduces fitness of both wildtype and resistant virus, then it will reduce the probability that a pre-existing resistant mutant becomes established. This is true even if the effect of the added drug on wildtype and resistant virus is exactly the same. Similarly, if the immune system works well, this may also reduce the probability of establishment. 

In most population genetics models, the focus is on the fixation probability, rather than the establishment probability of a mutation. And in many models, if a mutation becomes established, it will go to fixation. However, if selection pressures change, establishment does not necessarily lead to fixation. This is especially clear when we will later consider the effect of a single-dose of nevirapine. A few weeks after a single dose of nevirapine,  nevirapine resistance mutations can be detected in a large proportion of patients, but these mutations may never take over the whole viral population, because the treatment duration is very short and wildtype virus will quickly become a majority again (see for example, Lockman 2007). In fact, the standard results on fixation probability (Haldane 1927) are really results on establishment probabilities, so we can use them without problems.    

\textbf{Psgv vs. Pnew.}
For drug resistance to evolve, the viral population needs viral particles that carry drug resistance mutations. Such particles may already be present before treatment is started. To denote this possibility we use $P_{sgv}$ or the probability that drug resistance establishes from the standing genetic variation.  
If the mutation is not already present, or if was present but was subsequently lost, then the viral population has to wait for a new mutation to occur and become established. We denote this possibility as $P_{new}$, or the probability that resistance evolves due to new mutations. In the latter case, we have to indicate a time window, such as per year or per generation.  

The goal of this study is to understand and, albeit roughly, quantify $P_{new}$ and $P_{sgv}$ for HIV drug resistance in patients on triple-drug regimes (consisting of an NNRTI or an unboosted PI plus two NRTI's) and in patients who are treated with single dose nevirapine.

\section{Starting standard therapy}
\label{Start of therapy}

When a patient starts anti-retroviral therapy for the first time, the viral population in that patient will move from an equilibrium without drugs to an equilibrium with drugs. At the pre-treatment equilibrium, the viral population size will at its equilibrium level ($N_u$), and resistance mutations are expected to be at mutation-selection-drift equilibrium, where most mutations will be present at very low frequencies (see, e.g., Paredes 2010). Note that mutation-selection-drift equilibrium is reached quickly for mutations that are very costly to the virus. So even though it may take years for neutral diversity to reach an equilibrium level in an HIV patient (Kouyos et al 2011), important drug resistance mutations which are $5$ or $10\%$ less fit than the wildtype are expected to reach their (dynamic) equilibrium in weeks or months. 

Standard population genetic theory predicts that the average frequency of a resistance mutation is equal to the mutation rate ($\mu$, per viral particle and per replication) divided by the relative cost ($C_{rel}$) of the resistance mutation, though drift causes actual frequencies to vary greatly between different time points and between patients (see also Gadhamsetty et al 2010). Even though the average frequency is independent of the population size, in larger populations, it is more likely that DRMs are present and the absolute number of drug resistant particles will, naturally, be higher. When treatment starts, resistance mutations will confer a fitness benefit to the virus and they can (but are not guaranteed to) increase in frequency and become established. The probability that this happens depends on the number of resistant particles in the population and on the establishment probability of a mutation that is present in a single particle. In Hermisson and Pennings (2005) we derived formulas to calculate the probability that adaptation to a new environment happens from the standing genetic variation ($P_{sgv}$). We will use 
the approximate equation 8 in Hermisson and Pennings (2005): 
\begin{equation}
\label{P_sgv}
P_{sgv} \approx 1-(
1+\frac
{0.5\;P_{est}}{C_{rel}}
)
^{-2\mu N_u}
\end{equation}

It is also possible to use the the number of resistant particles in a patient ($B$) and the fitness of these copies (in the environment with drugs) to calculate the probability that a resistance mutation becomes established:  
\begin{equation}
\label{P_sgv2}
P_{sgv} = 1-(1-P_{est})^{B}
\end{equation}
where we use the probability that all copies of the resistance mutation die out to calculate the probability that at least one survives. The probability that resistance mutations become established increases with the number of copies of resistant virus and the probability that any one of these survives. 

\vspace{5 mm}

\textbf{Evolution of resistance during therapy.}
If resistance did not evolve from standing genetic variation, it may evolve due to new mutations. The probability that this happens in a given year will depend on the number of generations ($G$) in a year, the mutation rate ($\mu$), the effective population size during antiretroviral treatment ($N_{ART}$) and the establishment probability of a mutation ($P_{est}$). In principle, the establishment probability during therapy may not be the same as in the very beginning of therapy, for example because the number of available cells which a particle can infect could be different. However, throughout this paper we will assume that $P_{est}$ depends only on the kind of therapy and not on how long a patient has been treated. Using a poisson approximation, we find that the pre year probability that resistance evolves is   

\begin{equation}
\label{Rate}
P_{new} = 1-exp\;(-G \;N_{ART} \;\mu \; P_{est})
\end{equation}  

It is debated whether during therapy, there is ongoing replication or whether a reservoir of latently infected cells is entirely responsible to residual viremia. If the reservoir reflects the composition of the viral population before treatment, then the expected frequency of the resistance mutation in the reservoir would be $\frac{\mu}{C_{rel}}$. If the number of latently infected cells that become activated every generation is $N_{LAT}$, then the expected number of activated cells with resistant virus would be $\frac{N_{LAT}\;\mu}{C_{rel}}$. The per year probability that resistance evolves due to activated cells from the reservoir would be 

\begin{equation}
\label{Rate2}
P_{new} = 1-exp\;(\frac{-G \;N_{LAT} \;\mu \; P_{est}}{C_{rel}})
\end{equation}  

It is also possible that there is ongoing replication, but that the reservoir also plays a role at the same time, so that the reality will be reflected best by a combination of equations \ref{Rate} and \ref{Rate2}. 

\begin{figure2}[h]
\begin{center}
\includegraphics[width=7cm]{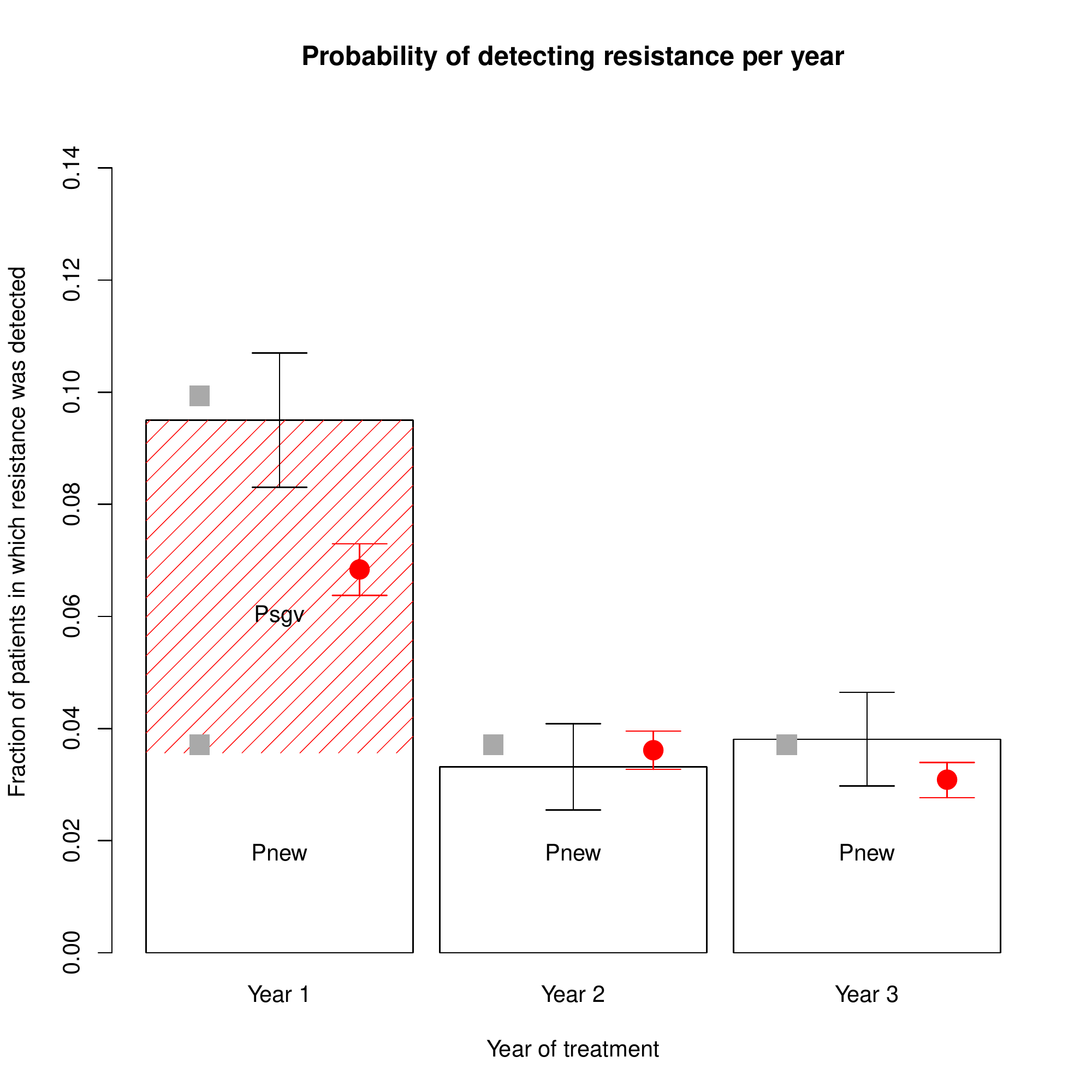}
\caption{The probability that resistance is detected for the first time in the first, second or third year of treatment, given that it was not detected until then. The bars are the estimates from the Margot et al (2006) dataset. The red dashed area reflects the inferred probability that resistance mutations from standing genetic variation become established. The grey squares are values calculated using equations \ref{P_sgv} and \ref{Rate}. The red circles are estimated from 1000 simulations. Parameters as in table \ref{TableParameters}.}
\label{Fig_Margot}
\end{center}
\end{figure2}

\begin{figure2}[h]
\begin{center}
\includegraphics[width=8cm]{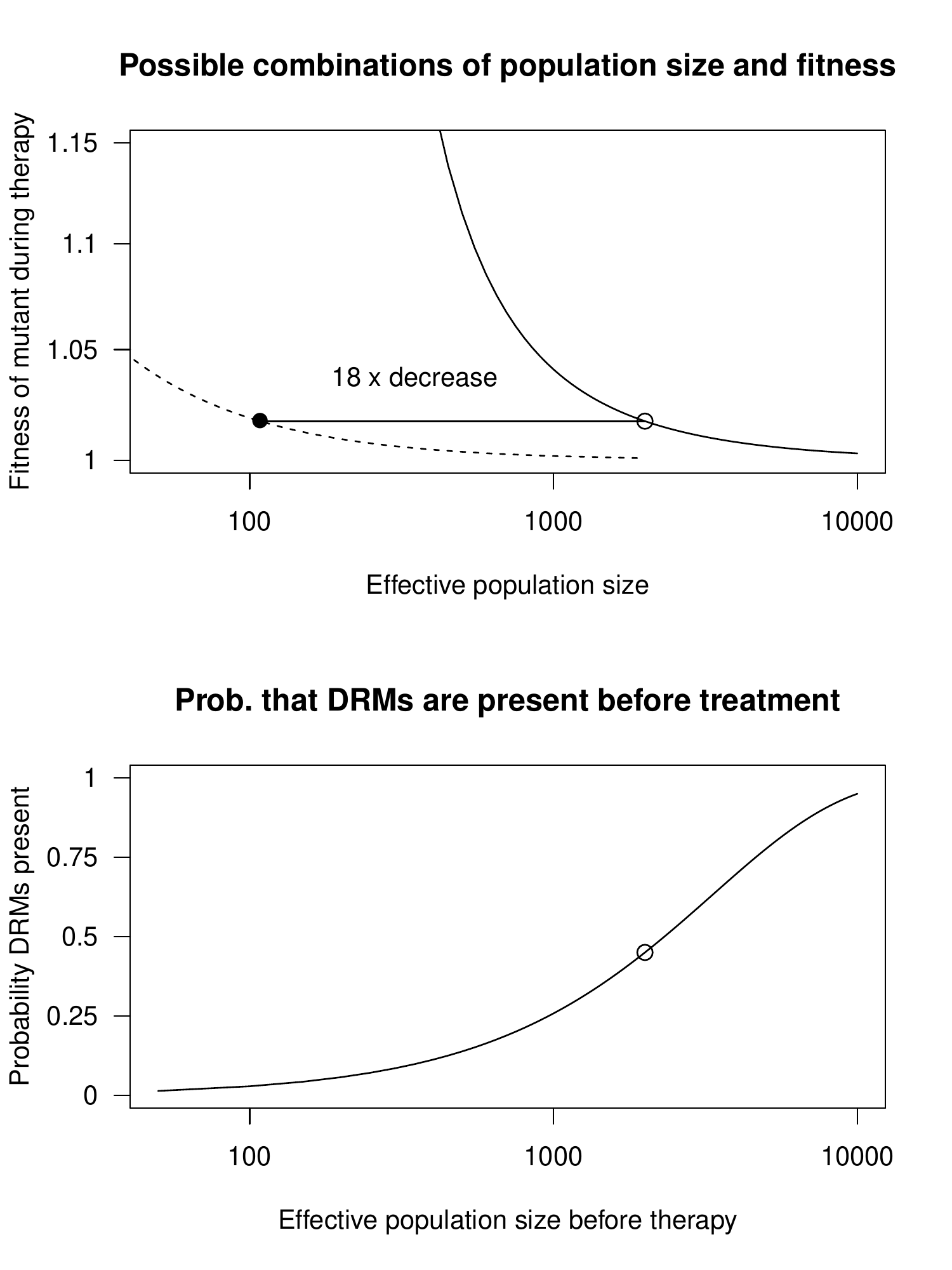}
\caption{2a. Continuous line: combinations of population size before treatment ($N_u$) and fitness of mutant virus during therapy ($Fm_{ART}$ that lead to the observed probability that resistance mutations from standing genetic variation become established ($P_{sgv}=0.058$). Dashed line: combinations of population size during treatment ($N_{ART}$) and fitness of mutant virus during therapy ($Fm_{ART}$ that lead to the observed probability that resistance mutations from standing genetic variation become established ($P_{sgv}=0.058$). Open dot: $N_u=2000$ and $Fm_{ART}=1.017$, closed dot: $N_{ART}=108$, $Fm_{ART}=1.017$. 
2b. Probability that a patient has any pre-existing DRMs before the start of therapy for different population sizes, and $\mu=5*10^-5$. Open dot: $N_u=2000$}%PARAM
\label{Combinations}
\end{center}
\end{figure2}

\textbf{Comparison with data and parameter estimation}
Published data show that the rate of evolution of drug resistance is roughly constant over long times (see for example the study by Cozzi-Lepri et al (2010), in which patients were followed for up to eight years). This fits with expectations if $N_{ART}$ and $P_{est}$ remain constant so that $P_{new}$ stays constant. However, several studies show that the probability that resistance mutations become established is higher in the first year of therapy, as compared to later years. This can be seen, for example, in a study on a large cohort in British Columbia, especially when one considers the most adherent group of patients (figure 2 in Harrigan et al 2005, see also Tam et al 2008). A similar effect is seen in Li et al (2011) when one considers the patients with pre-existing DRMs. This effect, that resistance is more likely to evolve in the first year of therapy as compared to later years, can be easily explained by standing genetic variation. 
 
Under the assumption that $P_{new}$ is indeed constant, we can use published data to estimate both $P_{new}$ and $P_{sgv}$. Margot et al (2006) reported the number of patients in which resistance was detected in the first, second and third year after treatment initiation in a cohort that was treated with NNRTI-based ART. The reported data show that the probability that resistance was detected in the first year was 9.5\%, whereas in the second and third year it was only 3.7\% (see supplementary material for details on how this was estimated). The difference of 5.8\% is likely due to standing genetic variation at the start of therapy. 

We will use the estimates for $P_{new}$  (0.037 per year) and $P_{sgv}$ (0.058) from Margot et al (2006), in combination with other, published, estimates to get a rough estimate of the important evolutionary parameters. First of all, we will assume that the mutation rate from one nucleotide to a specific other nucleotide is 
$10^{-5}$ (Mansky and Temin 1995)%PARAM
, so that if there are five main resistance mutations for a given drug combination, the total mutation rate is approximately 
$5\;10^{-5}$%PARAM
. For the remainder of the paper, we will only use this total mutation rate. If the mutation rate would be higher (lower) than our assumption, the estimated population sizes would be lower (higher) than our estimates. 

We know that the important drug resistance mutations are at least somewhat costly for the virus. Their cost, $C_{rel}$, has been estimated for several drug resistance mutations, both in vivo and in vitro (for an overview on resistance mutations in the reverse transcriptase gene see Martinez-Picado 2008). For example, Paredes et al (2009) find that the relative cost of resistance mutation M184V is approximately $0.04 - 0.08$. Wang et al (2011) estimate a cost of $0.01 - 0.04$ for K103N, which is the most common NNRTI resistance mutation. Other studies were not able to detect any cost of K103N, but given its low frequency in untreated patients (Paredes et al 2010), it seems likely that it is associated with a significant cost. In this paper we will use an average cost of 
$0.05 $ for all mutations. 
%PARAM
 
Given the cost, the mutation rate, $P_{sgv}$ and $P_{new}$, and using the assumption that there are 200 HIV generations in a year (Fu 2001), we can find the combinations of $N_u$, $N_{ART}$ and $Fm_{ART}$ that are compatible with the data (shown in figure \ref{Combinations}). 
Estimates for the effective population size  in untreated patients range from $10^3$ (Leigh-Brown 1997) to $10^5$ (Rouzine and Coffin 1999). 
We know that a large proportion of untreated patients carries low frequency drug resistance mutations, but not all patients, which gives us some additional information about the population size in an untreated patient (see figure \ref{Combinations}b). If we choose a value of $N_u$ of 
$2 \; 10^3$%PARAM
, then we find that about half of the patients should carry pre-existing DRMs. This is somewhat higher than what is usually detected, but that can be due in part to the limits of detection of current tests (Gianella and Richman 2010).  An overview of parameter estimates which we used in the simulations and for analytical predictions can be found in table \ref{TableParameters}.

Given our choice of $N_u$, we find that $Fm_{ART}$ must be approximately 
$1.017$%PARAM
, leading to 
$P_{est}\approx0.034$%PARAM
. Under the assumption $P_{est}$ stays the same during treatment, the Margot et al data are compatible with a 
18-fold %PARAM
 reduction of population size due to therapy, to a population size of 
 $N_{ART}\approx108$%PARAM
 . Note however, that the estimate of a 
 18-fold reduction %PARAM
 depends heavily on the assumption that 
 $C_{rel}=0.05$%PARAM
 . For example, if we had assumed a 10\% cost, the estimated reduction would have been 
 37-fold%PARAM
 , and for a 1\% cost, the reduction would have been only 
 4-fold%PARAM
 . The reason is that if we assume that costs are high, then we must also assume that the mutant fitness ($Fm_{ART}$) is relatively high, in order to find $P_{sgv}=0.06$, and if $Fm_{ART}$ is high, $N_{ART}$ must be low, to explain $P_{new}=0.037$. 
 
If the evolution of resistance during therapy is not due to ongoing replication, but due to continuous activation of latent cells, then, under the assumption that $C_{rel}=0.05$%PARAM
, the number of cells ($N_{LAT}$) must be approximately $5$ per generation. This means a reduction of population size of almost 400-fold. However, it is not so clear whether in this case the word ``population size" should still be used, because the number $5$ is not an estimate of the size of the reservoir, but an estimate of the size of the part of the reservoir that is reactivated every generation.
 
The result that the frequency of resistance mutations in the reservoir depends on their fitness cost ($\frac{\mu}{C_{rel}}$), whereas the cost does not play a role for new mutations due to ongoing replication, could be harnessed to estimate the relative importance of the reservoir. If the reservoir is the most important source of resistance mutations during therapy, then the same set of mutations should be found in patients whose virus acquires resistance quickly after the start of therapy and in those who acquire mutations during therapy. However, if ongoing replication is the source of resistance mutations during therapy, then mutations with a high cost in the absence of drugs should occur relatively more often during therapy than quickly after therapy is started. 
  
The data and the results from simulations and predictions (using equations \ref{P_sgv} and \ref{Rate}) are shown in figure \ref{Fig_Margot}. The percentage of patients with resistance after one year is lower in the simulations than in the analytical predictions, because it takes time for a mutation to increase in frequency and be detected. We assume that it is detected as soon as it is more frequent than the wildtype, the result is that in the simulations (and probably in reality) $P_{new}$ is lower in the first year than in the other two years.      

{\tiny
\begin{table}[h]
\caption{Parameter values for analytical predictions and computer simulations.}
\begin{center}
\begin{tabular}{|p{4.2cm}|p{1.2cm}|p{10cm}|}
\hline
Parameter&Value&Explanation \\ \hline

\textbf{Values roughly based on literature}\\ \hline

$\mu$&5*10-5&Mutation rate to resistant genotype\\ \hline
$N_u$&2000&Effective population size in untreated patient\\ \hline
$C_{rel}$&0.05&Relative cost of mutant in absence of therapy\\ \hline
$Fwt_{ART} = Fwt_{NVP}$&0.5&Absolute fitness of wildtype during therapy\\ \hline
$G$&200&Number of HIV generations per year\\ \hline

\textbf{Values estimated based on data}\\ \hline

$N_{ART}$&108&Effective population size in patient on ART\\ \hline
$N_{LAT}$&5&Number of activated latent cells in patient on ART\\ \hline
$N_{ZDV}$&1000&Effective population size in patient on ZDV monotherapy\\ \hline
$Fwt_{u}$&1.62&Absolute fitness of wildtype in absence of therapy 
(determines growth rate during treatment interruption)\\ \hline
$Fm_{u}$&1.54&Absolute fitness of resistant mutant in absence of therapy\\ \hline
$Fm_{ART}$&1.017&Absolute fitness of resistant mutant during ART\\ \hline
$Fm_{NVP}$&1.54&Absolute fitness of resistant mutant during NVP therapy\\ \hline
$Fm_{NVP/PP}$&1.025&Absolute fitness of resistant mutant during NVP/PP therapy\\ \hline
$I$&135&Number of particles from latent cells that enter the population per generation\\ \hline
\end{tabular}
\end{center}
\label{TableParameters}
\end{table}
}

{\tiny
\begin{table}[h]
\caption{Abbreviations.}
\begin{center}
\begin{tabular}{|p{2.9cm}|p{13cm}|}
\hline
Abbreviation&Explanation \\ \hline
VL& Viral load, the number of viral particles per ml blood\\ \hline
ART& Antiretroviral therapy, here used to mean treatment with two NRTIs and an NNRTI or an ``unboosted" PI\\ \hline
PMTCT&Prevention of mother to child transmission\\ \hline
DRM&drug resistance mutation\\ \hline
NRTI& Drug of class nucleoside reverse transcriptase inhibitor\\ \hline
NNRTI& Drug of class non-nucleoside reverse transcriptase inhibitor\\ \hline
PI& Drug of class protease inhibitor, PIs can be used ``unboosted" or ``boosted" with an additional drug.  \\ \hline
NVP& Nevirapine, an NNRTI\\ \hline
sdNVP& Single dose nevirapine\\ \hline
ZDV& Zidovudine, also known as AZT, an NRTI \\ \hline
3TC, DDI, FTC, TDF&Drugs of NRTI class\\ \hline
%&\\ \hline
%&\\ \hline
%&\\ \hline

\end{tabular}
\end{center}
\label{TableAbbreviations}
\end{table}
}

\section{Single-dose nevirapine for prevention of mother-to-child-transmission}\label{pregnant}
A single dose of nevirapine (sdNVP) just before labor starts reduces the risk that a mother transmits HIV to her baby at birth, but leads to high levels of resistance in many women. 
Because of the long half life of nevirapine, even a single dose lasts at least a few days. However, this is a very short amount of time (only a few HIV generations) so that probably most or all detected NVP resistance mutations are due to standing genetic variation. 

Because it is known that sdNVP can lead to the establishment of resistance mutations, and also to further reduce the risk that the baby becomes infected with HIV, several different treatment strategies are being used. In this study, we focus only on those strategies that include a single dose of nevirapine (and exclude, for example, pregnancy limited triple-drug therapy). Basically, sdNVP can be combined with either a short course of zidovudine monotherapy during the third trimester of pregnancy (ZDV/sdNVP), or it can be combined with additional drugs during and after labor up to one month postpartum (sdNVP/PP). It can also be used alone (sdNVP) or combined with both (ZDV/sdNVP/PP), resulting in four possible strategies. 

Under the assumption that all resistance is due to standing genetic variation, it is  straightforward to predict, at least qualitatively, the effect of the four treatment options.  
Single dose nevirapine plus two additional drugs (sdNVP/PP) is a three drug regimen, and similar to standard antiretroviral therapy (ART), except that it only lasts a few days or weeks. We therefore expect similar levels of drug resistance due to standing genetic variation. If only NVP resistance is considered (and not resistance to the other two drugs), we expect to find somewhat lower levels than in the normal case, although the difference may not be large because resistance against NVP is more common than resistance to most other drugs.  
Treating with only sdNVP is different from starting ART, in that there is only one drug. The result is that the fitness of both wildtype and resistant virus will not be reduced as much as in the normal case. Specifically, NVP resistant virus will have a relatively high fitness during NVP monotherapy. This high fitness  ($Fm_{NVP}$) leads to a high establishment probability ($P_{est}$) for available resistance mutations. In fact, the establishment probability may be so high that in virtually all patients that carry some NVP resistance before treatment, the resistant virus will increase in frequency during NVP treatment. 
 
An interesting treatment option is to start with a few weeks of ZDV monotherapy before treating with a single dose of nevirapine. The ZDV treatment will reduce the population size of the virus, $N_u$, so that the probability that NVP resistance is available and the copy number of such resistant mutants if they are available will be lower by the time the patient is treated with NVP. ZDV monotherapy ultimately leads to ZDV resistance, but the risk that resistance mutations become established during a short course is small. ZDV monotherapy reduces the viral load approximately three-fold (BrunVezinet 1997). 
Finally, adding ZDV treatment before labor and two additional drugs during and after labor (ZDV/sdNVP/PP) will reduce both the availability of NVP resistant virus and the establishment probability of such virus, which should lead to an even lower probability that NVP resistance mutations from standing genetic variation become established. 

\begin{figure2}[h]
\begin{center}
\includegraphics[width=10cm]{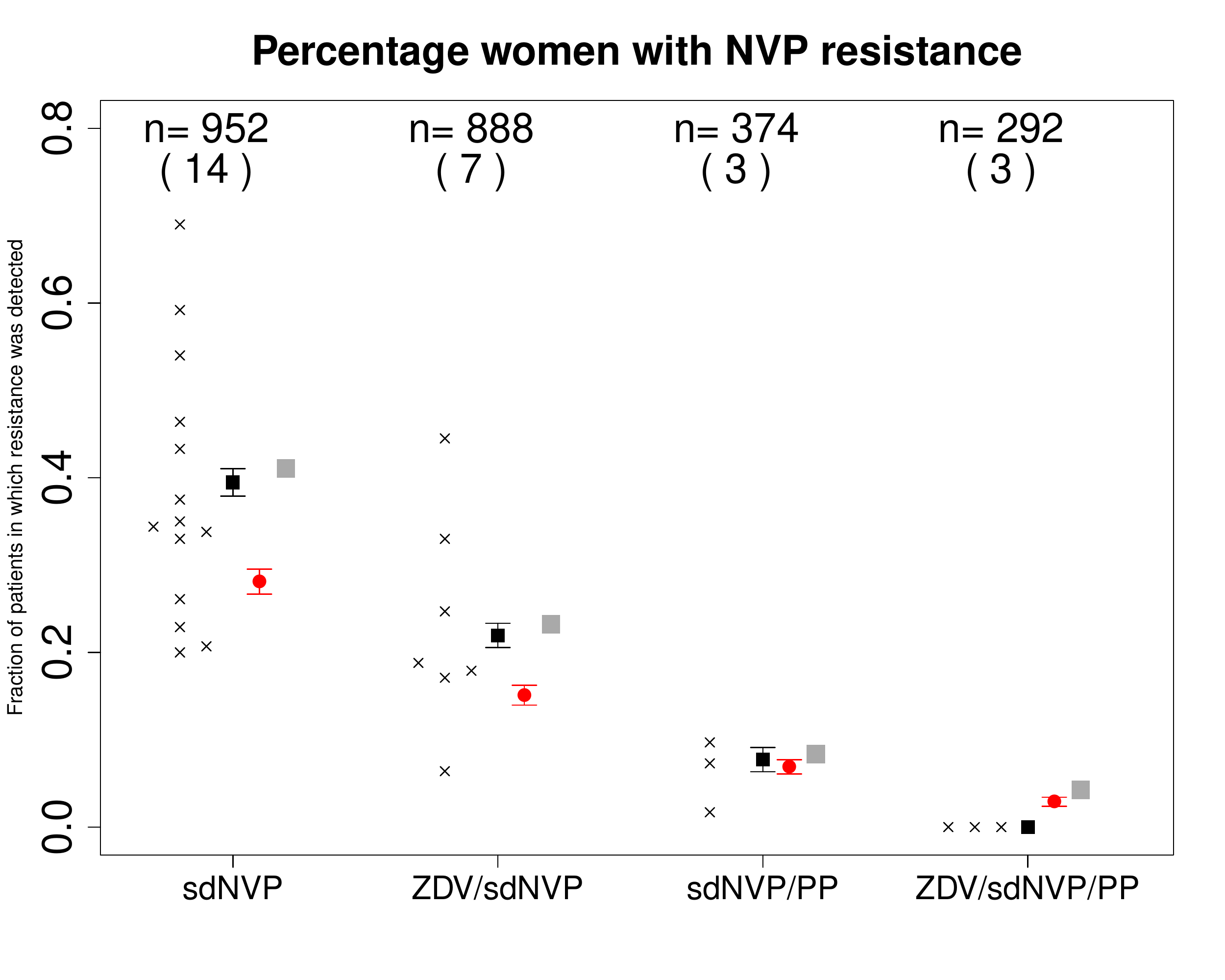}
\caption{The probability that resistance mutations are detected 6 to 8 weeks after treatment with single dose nevirapine. Black crosses are data from single studies (the number of studies is indicated in brackets at the top of the graph), black squares with estimated standard error are percentages for all studies combined (the number of patients that were used to calculate this percentage is indicated at the top of the graph). Red circles with standard error are results from 1000 simulations and the grey squares are analytical predictions. Parameter values as in table \ref{TableParameters}. sdNVP stands for single dose nevirapine, ZDV stands for third trimester zidovudine monotherapy, PP stands for the addition of two drugs during and after labor (postpartum).}
\label{PMTCT}
\end{center}
\end{figure2}

\textbf{Comparison with data for single dose nevirapine.}
We identified 23 published studies that reported on NVP resistance 6 to 8 weeks after women were treated with sdNVP. Several of the studies directly compared two different treatment options. We found at least three studies for each of the four different treatment options.  An overview of the studies can be found in table \ref{TableNVP} in the supplementary material. For each study we recorded which of the four treatment options was used and in how many of the patients NVP resistance mutations were detected using simple Sanger (population) sequencing (we excluded studies that only recorded deep-sequencing or allele-specific PCR results, as there were too few of those to allow us to compare the treatment options). For each of the four treatment options, we also calculated the overall probability that resistance mutations were detected in a patient (simply by summing the number of patients with resistance and summing the total number of patients in the studies). We found that sdNVP leads to detectable resistance mutations in 39\% of 952 patients, ZDV/sdNVP leads to detectable resistance mutations in 22\% of 888 patients, adding two drugs during and after labor (sdNVP/PP) lead to detectable resistance mutations in 7.8\% of 372 patients and ZDV/sdNVP/PP lead to detectable resistance mutations in none of 292 patients. 
 
We now used these data, in combination with our previous parameter estimates, to estimate the fitness of a NVP resistant mutant during NVP therapy ($Fm_{NVP}$) and the reduction of the population size due to ZDV treatment ($N_{ZDV}$). We find that $Fm_{NVP}\approx 1.54$ and that ZDV reduces the effective population size approximately two-fold (table \ref{TableParameters} and figure \ref{PMTCT}). 
The results show that a reduction in population size by ZDV monotherapy does reduce the probability that NVP resistance mutations become established, but adding two drugs to sdNVP helps much more. We also estimate of the fitness of the mutant during therapy with nevirapine and two additional drugs and find a slightly higher value than our previous estimate ($1.025$ vs $1.017$), because the estimated probability that resistance mutations from standing genetic variation become established is slightly higher ($0.078$ vs $0.058$ with standard therapy), though these differences are not statistically significant. 

\begin{figure2}[h]
\begin{center}
\includegraphics[width=8cm]{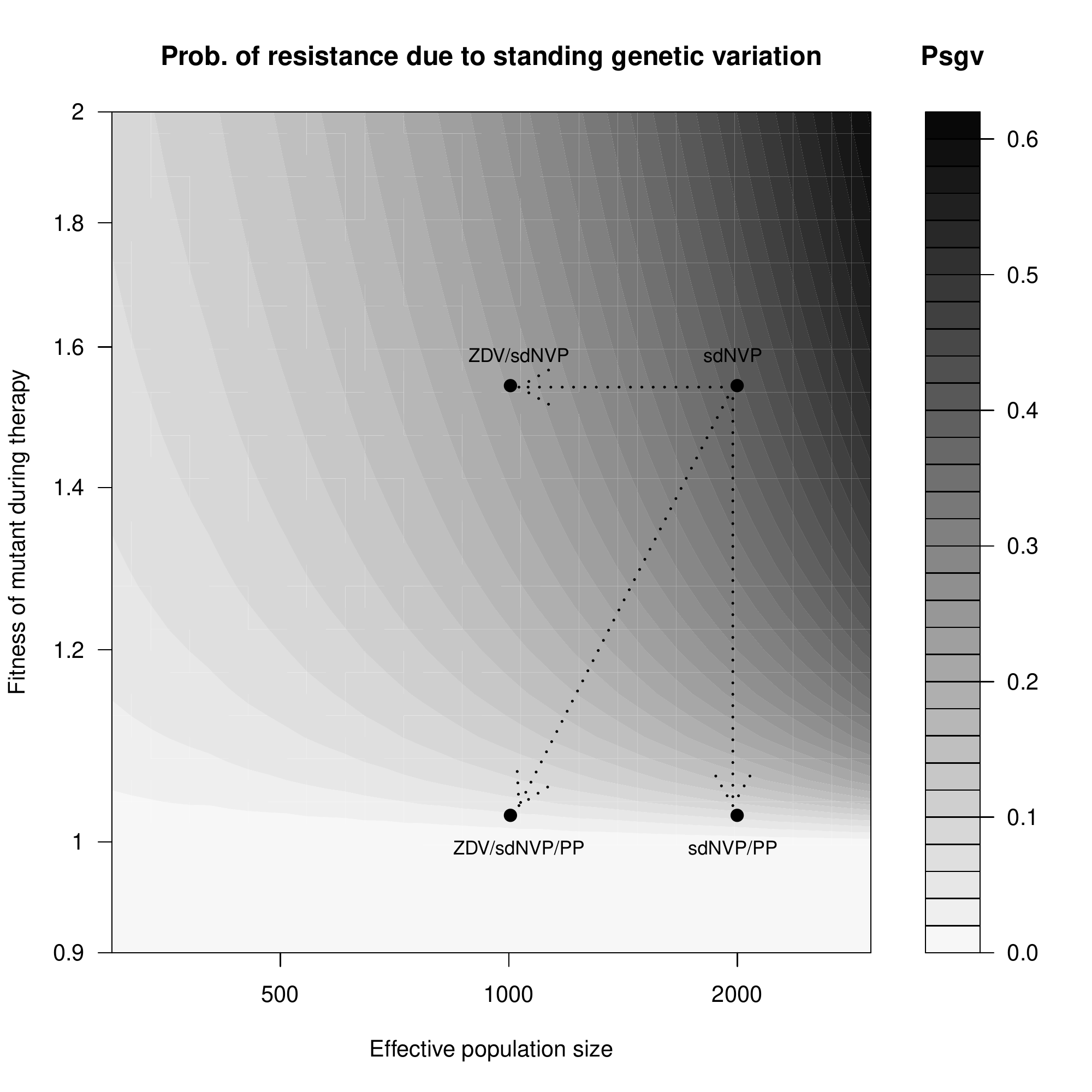}
\caption{The probability of the establishment of drug resistance mutations from standing genetic variation depending on the effective population size and the fitness of the resistant mutant during therapy. Grey scales indicate the probability of the evolution of drug resistance due to standing genetic variation. Dots indicate estimated parameter combinations for treatment with just sdNVP, with ZDV monotherapy followed by sdNVP (ZDV/sdNVP), by sdNVP followed by two additional drugs postpartum (sdNVP/PP) and with ZDV monotherapy followed by sdNVP and two additional drugs postpartum ZDV/sdNVP/PP. }
\label{NVPCombinations}
\end{center}
\end{figure2}

\section{Interruption of therapy}
\label{Interruption of therapy}

During a treatment interruption, drugs are first removed from the body, which can take from a couple of hours to a several days  or even weeks (Van Heeswijk 2000, 2001, Pirillo 2010). With some delay, depending on the half-life of the drugs, the viral population begins to grow, which is observed as an increase of viral load. 
Published data show that after treatment is stopped, viral load quickly increases in almost all patients (e.g., Harrigan 1999). Davey et al (1999) show that average viral load plateaus four weeks after treatment is interrupted. Garcia et al (2001) and Trkola et al (2005) both report that a plateau is reached between four and eight weeks after treatment interruptions. An interruption is ended when treatment is started again and viral load goes down, hopefully to undetectable levels. Figure 1 shows a cartoon of the pharmacodynamics and population dynamics of a treatment interruption. 

\begin{figure2}[h]
\begin{center}
\includegraphics[width=7cm]{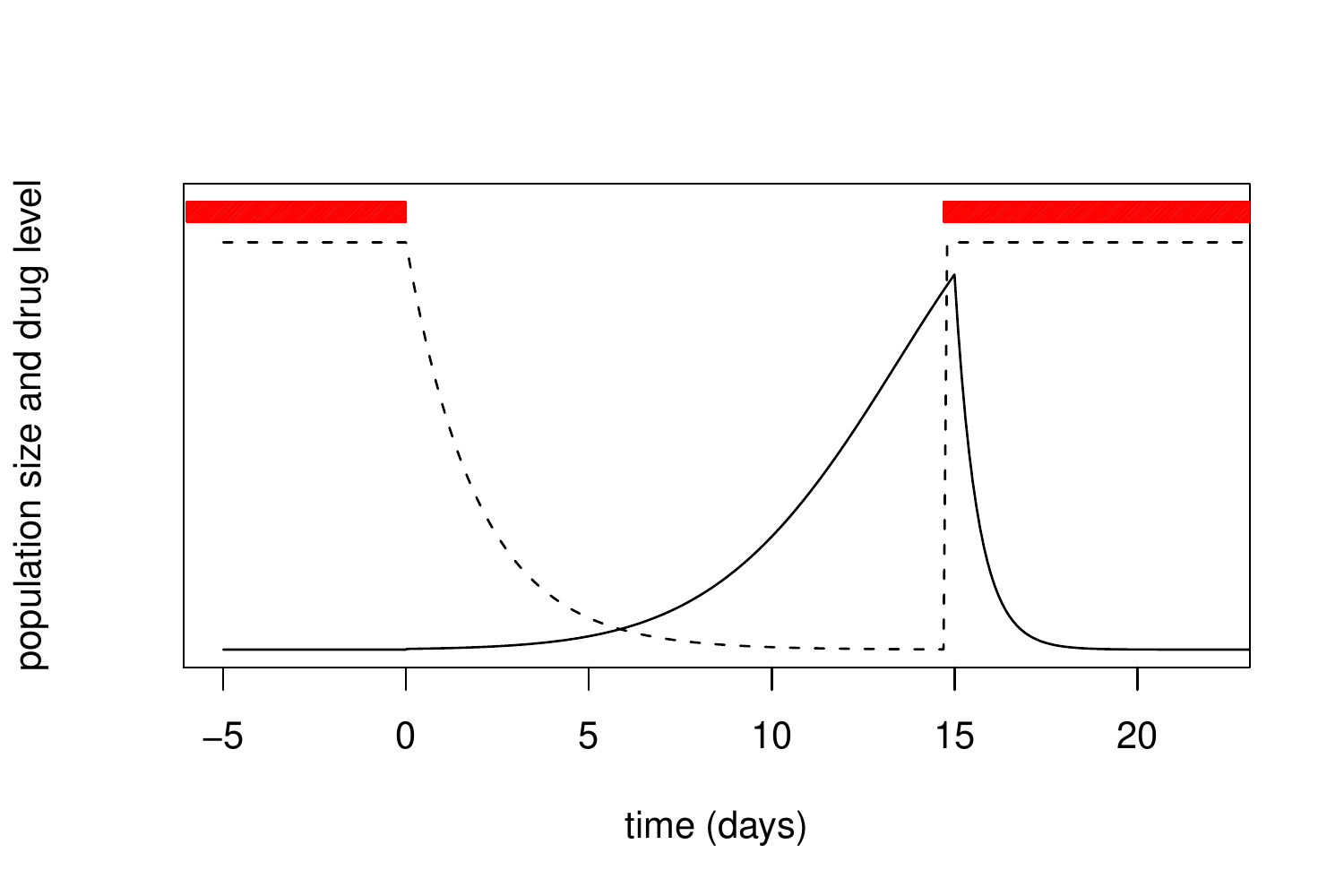}
\caption{Drug level  (dashed line) and viral population size (solid line) during and after a treatment interruption. Red bars indicate when drugs are taken.}
\label{TimeScales}
\end{center}
\end{figure2}

\textbf{Restarting therapy.}
If the length of a treatment interruption is so long that the population size is back to pretreatment level and mutation-selection-drift equilibrium is again reached, the probability that resistance mutations become established when therapy is started again will equal the probability that resistance mutations become established the first time a patient starts treatment, $P_{sgv}$ from equation \ref{P_sgv}. But if a treatment interruption is shorter than that, it is hard to calculate the exact probability that resistance will evolve upon re-initiation of therapy because neither population-dynamic, nor population-genetic equilibrium will have been reached. The absence of the population-genetic equilibrium is most problematic if resistance mutations are not very costly to the virus. However, for a costly mutation it takes only on the order of $1/C_{rel}$ generations to reach mutation-selection-drift equilibrium.   
The absence of population-dynamic equilibrium is less problematic, because it is relatively easy to predict the population size of the virus or to measure viral load. In the simulations, we allow the population to grow exponentially until it reaches the baseline level. The resulting population size can be plugged into equation \ref{P_sgv} to get an estimate of the probability that resistance mutations become established due to a treatment interruption. 

\textbf{Comparison with data for treatment interruptions.}
Using the parameter values from the last two sections, we can predict the risk that resistance mutations become established due to a treatment interruption of a certain length. We use the estimated fitness of the mutant virus during NVP therapy, and assume that the fitness of the mutant in absence of drugs is the same. With that value, we can calculate the fitness of the wildtype in the absence of drugs, because of the assumption that the cost of the resistance mutation is 5\%. The wildtype fitness will determine how fast the virus grows in the simulations after treatment is interrupted, and therefore how long it takes before the population size is back at the pretreatment level. Specifically, we use $F_{wt}=1.62$. In the simulations, the population size plateaus after just 14 days, but $P_{sgv}$ reaches its expected value only after 60 days (figure \ref{STIData}). 

We collected information from structured treatment interruption trials to test the predictions. 
The probability that resistance mutations become established due to a single treatment interruption was estimated for seven clinical trials with different lengths of treatment interruptions (Ananworanich, 2003, Danel, 2009, Hoen, 2005, Palmisano, 2007, Reynolds, 2009 and 2010, Yerly 2003). An overview of the trials can be found in table \label{TableSTI}
 (supplementary material). We first calculated the risk under the assumption that all observed resistance was due to treatment interruptions and then subtracted the estimated probability that resistance mutations become established during therapy. The corrected values are shown in figure \ref{STIData}. The data show that longer treatment interruptions indeed lead to a higher risk of resistance. The risk plateaus around 37 days, which is consistent with the time it takes for viral load to reach its equilibrium level (although the simulations suggest that the risk should plateau later than the population size). The highest risk was found to be approximately 6\% per interruption, just like the risk of starting therapy for the first time. 

\begin{figure2}[h]
\begin{center}
\includegraphics[height=6cm]{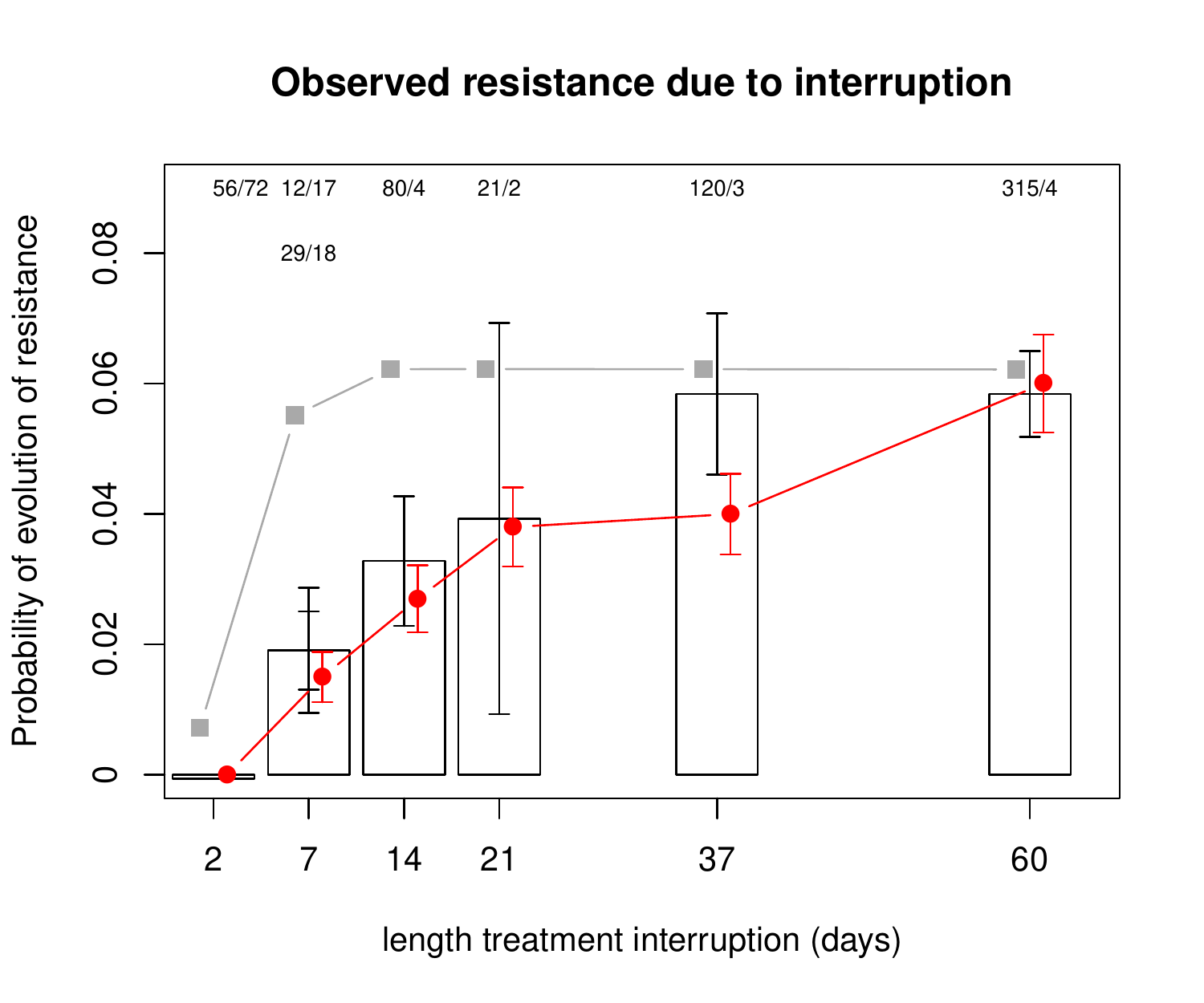}
\caption{Estimated probability that resistance mutations become established due to a single treatment interruption. Bars are data from seven clinical trials, $+/-$ estimated standard error (see table \ref{TableSTI}, supplementary material), 
the red circles are estimated from 1000 simulations, $+/-$ estimated standard error
the grey squares are predictions using the average population size from the simulations and equation \ref{P_sgv}.
Parameters as in table \ref{TableParameters}.}
\label{STIData}
\end{center}
\end{figure2}

\section{Discussion}

One aim of our study was to understand and quantify the importance of standing genetic variation for the evolution of drug resistance in HIV. We find that the probability that at least one resistance mutation becomes established due to standing genetic variation ($P_{sgv}$) depends on the kind of treatment chosen. Most clearly, it is much higher when treatment is with sdNVP (which is monotherapy) than if treatment is with triple-drug combination therapy. For standard combination therapy (ART), we use two different data sources to estimate the probability that resistance mutations from standing genetic variation become established. In the first part of this paper we used data on the number of patients in which resistance was detected in the first year of treatment versus later years. In the third part of this paper we used data from clinical trials on treatment interruptions. In both cases, we found that the probability that resistance mutations from standing genetic variation became established was approximately 6\%. 

The importance of new mutations as compared to pre-existing mutations could be estimated from the Margot et al 2006 study. We estimated that the probability that a resistance mutation becomes established during therapy ($P_{new}$) is 3.7\% per year, which means that pre-existing mutations and new mutations are equally important after about one-and-a-half year of treatment. Two of the interruption studies also provided estimates for $P_{new}$, which were slightly higher (4.3\% and 4.8\% per year) than the estimate from the Margot et al 2006 study (see table \ref{TableSTI}).
It is likely that some of the patients in these studies were not perfectly adherent to treatment, so that our estimate of $P_{new}$ is inflated by patients who interrupted treatment. This does not affect our estimates of $P_{sgv}$. However, it means that the relative importance of pre-existing mutations is highest in completely adherent patients (because new mutations are relatively unimportant for them) and lower in non-adherent patients (see Paredes 2010, but see Li 2011). 

A stochastic model was used to understand the effect of standing genetic variation on the evolution of drug resistance during HIV treatment. Four parameters are crucial to understand the role of standing genetic variation. Three of them determine the amount of genetic variation that is available (effective population size, mutation rate and cost of the resistance mutations) and one determines how likely it is that the available mutations become established (the absolute fitness of the resistant virus during treatment). 

The cost and the mutation rate are parameters that are specific for each specific mutation. Together, they determine the expected frequency of the mutant in an untreated patient. 
For example, in untreated patients the frequency of K103N was found to be lower than the frequency of Y181C (Paredes et al 2010), suggesting that $\mu/C_{rel}$ is lower for K103N, possibly because it reduces the fitness of the virus more. 
The costs for some of the most important mutations (M184V, K103N) have been estimated and are between 1 and 10\% (Martinez-Picado 2008, Paredes 2009,  Wang 2011). Throughout this paper we used a value of 5\%.

The effective population size in an untreated patient ($N_u$) determines how much variation there is in the frequency of resistant mutants between patients. If $N_u>1/\mu$, the frequency in each patient will be very close to the expectation, $\mu/C_{rel}$, but if $N_u<1/\mu$, there will be a lot of variation between patients, and in many patients no resistance mutations may be available at all. Data suggest that in HIV the latter is the case (e.g., Paredes et al 2010), which means that, contrary to common belief, not every single point mutation is created every generation in an HIV patient. $N_u$ also determines the number of resistant viral particles in a patient with a given frequency of the mutant. With higher $N_u$, there will be a higher number of resistant particles, and this makes it more likely that resistance mutations become established when treatment is started (Li et al 2011). 

We find that data are compatible with an 18-fold reduction of $N$ due to ART and a two-fold reduction of $N$ due to ZDV monotherapy. The estimated reduction depends on the assumed cost of mutations; if we assume that mutations are twice as costly, we would find a reduction that is twice as severe. Still, the reductions we find are not nearly as severe as one may have expected based on viral load reductions. During ART, VL may be reduced 1000-fold or more (in the Margot (2006) from which we used the data study, patients had a viral load of, on average, $8\;10^4$ before treatment, whereas after 48 weeks of treatment, about 80\% of the patients had a viral load of less than 50, Gallant 2004). This discrepancy may be due to two effects: firstly, our estimate is an average for many patients and this average may be driven up by patients in which the drugs do not work well, or who are not adherent to therapy so that their VL does not go down as much as expected. Secondly, the relationship between effective population size and viral load may not be linear, so that a thousand-fold reduction in VL may translate in only a twenty-fold reduction in effective population size.  

The fourth important parameter is the fitness of the mutant virus during treatment ($Fm$), which determines the establishment probability ($P_{est}$). $Fm$ will depend on both the drugs that are used and on the specific mutation. For example, the resistance mutation K103N is more likely to become established during sdNVP than during triple-drug therapy, because additional drugs reduce $Fm$ ($Fm_{ART}<Fm_{NVP}$). And during triple-drug therapy, K103N is more likely to become established than Y181C (even though Y181C is present at higher frequencies before treatment), likely because $Fm_{ART}$ is higher for K103N than for Y181C. 

\textbf{Starting of standard therapy.}
We assumed that the rate of evolution due to new mutations is constant and that the establishment of a resistance mutation from standing genetic variation leads to viral failure and is detected within one year of starting therapy. Maybe the most convincing evidence for these assumptions comes from the Li et al (2011) study, where their figure 2 shows that (1) patients without detected pre-existing DRMs show a constant rate of evolution of resistance and (2) patients with detected pre-existing DRMs show an increased rate compared to the patients without pre-existing DRMs, but only in the first year of treatment. We used these assumptions to estimate the probability that resistance mutations from standing genetic variation become established. However, the estimated role of standing genetic variation may be a slight underestimate, because establishment of new mutations should need some time so that $P_{new}$ would normally be somewhat lower in the first year of treatment. The observation that the effect of standing genetic variation only lasts a year, means that fixation of a resistance mutation must take less than a year. This limits possible values for $N_{ART}$ and $Fm_{ART}$ to such values for which the fixation time is less than 200 generations. For the current study, this means that, had we started with a larger population size estimate, then we would have had to choose a higher value of $Fm_{ART}$ in order to guarantee that pre-existing resistance mutations are detected (or lost) within a year of therapy, which, in turn, would have led to a higher value of $C_{rel}$ to keep $P_{sgv}\approx6\%$.  

If resistance indeed evolves due to standing genetic variation in 6\% of patients on standard ART, then there is clearly room for improvement. Note that those 6\% of patients have already lost their first treatment option shortly after having started treatment. They have to switch to second-line treatment which is more expensive, usually more complicated (more pills per day) and likely has more side effects. It is therefore worth exploring ways to avoid the establishment of resistance mutations from standing genetic variation. Figure \ref{NVPCombinations} suggests two options to reduce $P_{sgv}$, by reducing the population size or by reducing the fitness of the resistant mutants. The first may be achieved by ZDV monotherapy, as shown in the section on PMTCT, whereas the second may be achieved by adding additional drugs to the treatment. Obviously, triple-drug combination treatment is already standard for most HIV patients, but it may be worth considering specifically which treatment options would be best to prevent the evolution of resistance from standing genetic variation. This may mean, for example, to add a fourth drug to the therapy in the first couple of weeks of treatment. Finally, when it becomes possible to detect low-frequency DRMs routinely in patients before starting treatment, it may be possible to wait for a moment when their frequency is low to start therapy.  
  
\textbf{Resistance due to sdNVP. }
Studying treatment with a single dose of nevirapine gives us a unique opportunity to study the effect of standing genetic variation, because treatment is so short (only a few HIV generations) that we can assume that most or all resistance mutations that are detected are from standing genetic variation. Data show that the risk that resistance mutations become established due to such treatment is very high (39\%). We find that this high probability can be explained entirely by selection on pre-existing drug resistance mutations, because the fitness of NVP resistant virus is probably very high during NVP monotherapy. We estimate that its fitness is approximately 1.5. Reducing this fitness, by adding additional drugs or reducing the availability of the mutant by reducing population size will both reduce the probability that resistance is established. A study from Zambia (Chi et al 2007) showed that the additional drugs even help to reduce the establishment of NVP resistance mutations if the drugs are given as a single dose (in stead of treatment for a couple of days or weeks). We did not include this study in the overview, because there was only one study that looked at this treatment option. However, the results are interesting: they find that even a single dose of  TDF/FTC reduces the probability that resistance mutations are found 6 weeks after a sdNVP from 25 to 12 \%. 

The results on ZDV/sdNVP/PP treatment (i.e., treatment with ZDV during pregnancy and NVP plus two other drugs during labor) are surprising in that NVP resistance mutations were not detected in any of the women who received this treatment, even though the model would predict that mutations would be detected in 4\% of the women. Most of the data on this treatment option are from the Lallemant 2010 paper (222 women). In this  study, the authors do find some mutations that confer resistance to the NRTI's in the study (in 2.3\% of the women). The same study also looked at women who were treated with ZDV/sdNVP and also in these women the percentage with resistance mutations was very low (6.4\%) and much lower than the mean value for women who receive this treatment (22\%). The reason for the surprisingly low values of drug resistance in this study could be that the women in the study had very low viral loads (median 2800), which is much lower than in most untreated patients. This probably also means that they have a low effective population size. It therefore seems unlikely that the extremely good results from the Lallemant study can be replicated in other populations. However their results still show that using additional drugs to reduce the population size and to reduce the fitness of the mutant may be a good strategy to reduce the probability that resistance becomes established. 

\textbf{Treatment interruptions.}
Considering treatment interruptions, our model provides several testable predictions. 
1) resistance mutations are more likely to become established after long treatment interruptions when viral loads are higher, 
2) the risk that resistance mutations become established due to a treatment interruption can not be larger than the risk at the start of treatment,  
3) treatment interruptions increase the risk of establishment of resistance mutations even for drugs with short half-lifes. 

Data from seven clinical trials show that indeed, longer interruptions increase the probability that resistance mutations become established (figure \ref{STIData}). Moreover, the estimated probability appears to plateau after 37 days, which is similar to the time it takes for viral load to reach its pretreatment level. This suggests that the risk of establishment of resistance mutations is directly linked to the viral load when treatment is started again. However, in our simulations, population size plateaued much earlier than the risk of resistance. Unfortunately, the resolution of both viral load measurements and of the data on treatment interruptions are not good enough to determine whether in reality viral load indeed plateaus before the risk or not. If both would plateau at the same time, it would suggest that the resistance mutations are more costly than we had assumed. 
The second prediction was also found to hold: the estimated risk that resistance mutations from standing genetic variation become established at the start of treatment was found to be similar to the risk due to a long treatment interruption (6\% in both cases). 
The third prediction also holds, as data show that interruptions increase the risk of establishment of resistance mutations even for PI based treatment (Yerly 2003 and Hoen 2005), where the ``tail of monotherapy" cannot explain the observations. 

A potential problem with the data is that not only the length of the interruptions, but also the length of treatment periods between the interruptions differed between the seven studies. The trials that were compared also differed in the drugs that were used (see table \ref{TableSTI} in supplementary material), which makes direct comparison difficult. Despite all these limitations, it becomes clear that longer interruptions carry a higher risk of evolution of resistance than shorter interruptions. 

If interruptions lead to the establishment of resistance mutations only due to the ``tail of monotherapy", as is usually assumed in the HIV literature (Taylor 2007, Fox 2008, Gardner 2010), we would predict that: 
4) treatment interruptions increase the risk that resistance mutations become established only for drugs with long half-lifes, 
5) the risk that resistance mutations become established due to a treatment interruption is unrelated to the risk at the start of treatment and 
6) the largest risk would be due to an interruption with a length that is exactly the time it takes for the last drug to loose its effect on the wildtype virus. All of these predictions do not hold. This is not to say that the ``tail of monotherapy" is not important at all. But it does show that on its own, the ``tail of monotherapy" cannot explain the risk that resistance mutations become established due to treatment interruptions. When one considers possible intervention strategies, this may be good news. If treatment interruptions are risky because of restarting rather than stopping therapy, this would give doctors a possibility to reduce the risk that resistance mutations become established even after a patient has already stopped taking his or her drugs.  

The data on single dose nevirapine clearly show that monotherapy makes it likely that resistance mutations become established, so intuitively, one would expect that monotherapy should also be important also in patients who interrupt treatment, which is why the ``tail of monotherapy" hypothesis is so popular. However, in patients who are treated with sdNVP, the resistance mutations are most likely already present, because it is the first treatment they receive and the viral population will be large. On the other hand, if a fully suppressed patient interrupts therapy, a few days of monotherapy may not have a large effect, because the population size of the virus will be very small. Monotherapy in those patients would still lead to a high establishment probability of resistance mutations, but if such mutations are not present in the population, they cannot become established. The situation may be different in patients who take treatment very irregularly, so that they are more likely to have higher viral loads when treatment is stopped. This may explain why resistance to NNRTIs, which are long half-life drugs, is likely to evolve in non-adherent patients (Tam et al 2008).

\textbf{General remarks. }
We  have used a population-dynamic and population-genetic model to study several patterns of drug resistance in HIV. The model explains why resistance mutations are likely to be established in the first year of standard treatment, in women who are treated with a single dose of nevirapine and in patients who interrupt treatment. In all three cases, standing genetic variation can explain the observations. 

Our results illustrate that for adaptive evolution to happen, selection and the creation of new variation need not happen at the same time, if selection can work on standing genetic variation. In the case of antiretroviral treatment, this means that insufficient drug levels (which allow for replication and selection at the same time) are not a necessary condition for the evolution of drug resistance, although they would speed up such evolution. This result about time-heterogeneous drug levels is similar to the result on heterogeneity in space by Kepler and Perelson (1998), who showed that genetic variation may be created in compartments where drugs cannot penetrate whereas selection happens in other compartments. 

Note that our model provides a simple explanation for why resistance is less likely to evolve when patients are treated with multiple drugs at the same time in stead of just one drug. Additional drugs reduce the fitness of a mutant that is resistant against one drug, and therefore the establishment probability of such a resistant mutant. In addition, additional drugs reduce the population size of the virus and thereby the creation of new resistance mutations. This means that there will be fewer resistance mutations with lower establishment probabilities, together leading to a strong reduction in the probability that resistance evolves. This model therefore explains why the evolution of drug resistance is slower with triple-drug therapy, but not completely halted, as escape due to one mutation is still possible. In newer therapies with boosted PIs, drug resistance has become very rare (Cozzi-Lepri 2010), which may be because boosted PIs are so strong that no single mutation can lift the virus' fitness above 1.  

The model in this study, which includes the effect of population size on the probability that resistance mutations become established may be relevant to other diseases than HIV. 
For example, the evolution of resistance is a problem in chronic myeloid leukemia (CML) which is a cancer of white blood cells. A recent study suggested that the probability that drug resistance evolves in CML goes down with time because the population size of the cancer goes down with time (Tomasetti, 2011, BCJ).  

Resistance is also a problem in tuberculosis (TB), and in TB it is also known that treatment interruptions increase the risk of evolution of resistance (Weis 1994). It should be investigated whether this effect may also be due to an increased population size during the interruptions. In general, stopping treatment may be risky when treatment has to be started again, which is always the case for HIV and often for TB. Each time therapy is started, resistance mutations from standing genetic variation may become established, and even if this risk is only a few percent it adds up quickly when patients interrupt treatment regularly. 

\section{Acknowledgements}
This work was supported by a long-term fellowship of the Human Frontier Science Program. I would like to thank
Marijn van Ballengooijen,  
Sebastian Bonhoeffer, 
John Coffin,
Andreas Gros,
Joachim Hermisson,
Sergey Kryazhimskiy,
Dan Kuritzkes,
Rebecca Meredith,
Igor Rouzine,
George Shirreff,
John Wakeley
and
Meike Wittmann 
for fruitful discussions.

\section{Bibliography}

Ananworanich, J., R. Nuesch, et al. (2003). "Failures of 1 week on, 1 week off antiretroviral therapies in a randomized trial." Aids 17(15): F33-F37.

Arnedo-Valero, M., F. Garcia, et al. (2005). "Risk of selecting de novo drug-resistance mutations during structured treatment interruptions in patients with chronic HIV infection." Clinical Infectious Diseases 41(6): 883-890.

Arrive, E., M. L. Chaix, et al. (2010). "Maternal and nenonatal tenofovir and emtricitabine to prevent vertical transmission of HIV-1: tolerance and resistance." Aids 24(16): 2478-2485.

Arrive, E., M. L. Newell, et al. (2007). "Prevalence of resistance to nevirapine in mothers and children after single-dose exposure to prevent vertical transmission of HIV-1: a meta-analysis." International Journal of Epidemiology 36(5): 1009-1021.

Barrett, R. D. H. and D. Schluter (2008). "Adaptation from standing genetic variation." Trends in Ecology \& Evolution 23(1): 38-44.

Bonhoeffer, S. and M. A. Nowak (1997). "Pre-existence and emergence of drug resistance in HIV-1 infection." Proceedings of the Royal Society of London Series B-Biological Sciences 264(1382): 631-637.

Brown, A. J. L. (1997). "Analysis of HIV-1 env gene sequences reveals evidence for a low effective number in the viral population." Proceedings of the National Academy of Sciences of the United States of America 94(5): 1862-1865.

BrunVezinet, F., C. Boucher, et al. (1997). "HIV-1 viral load, phenotype, and resistance in a subset of drug-naive participants from the Delta trial." Lancet 350(9083): 983-990.

Chaix, M. L., D. K. Ekouevi, et al. (2007). "Impact of nevirapine (NVP) plasma concentration on selection of resistant virus in mothers who received single-dose NVP to prevent perinatal human immunodeficiency virus type 1 transmission and persistence of resistant virus in their infected children." Antimicrobial Agents and Chemotherapy 51(3): 896-901.

Chalermchockcharoenkit, A., M. Culnane, et al. (2009). "Antiretroviral Resistance Patterns and HIV-1 Subtype in Mother-Infant Pairs after the Administration of Combination Short-Course Zidovudine plus Single-Dose Nevirapine for the Prevention of Mother-to-Child Transmission of HIV." Clinical Infectious Diseases 49(2): 299-305.

Chi, B. H., G. M. Ellis, et al. (2009). "Intrapartum Tenofovir and Emtricitabine Reduces Low-Concentration Drug Resistance Selected by Single-Dose Nevirapine for Perinatal HIV Prevention." Aids Research and Human Retroviruses 25(11): 1099-1106.

Chi, B. H., M. Sinkala, et al. (2007). "Single-dose tenofovir and emtricitabine for reduction of viral resistance to non-nucleoside reverse transcriptase inhibitor drugs in women given intrapartum nevirapine for perinatal HIV prevention: an open-label randomised trial." Lancet 370(9600): 1698-1705.

Cozzi-Lepri, A., D. Dunn, et al. (2010). "Long-Term Probability of Detecting Drug-Resistant HIV in Treatment-Naive Patients Initiating Combination Antiretroviral Therapy." Clinical Infectious Diseases 50(9): 1275-1285.

Dabis, F. and T. E. A. S. Grp (2009). "Tolerance and viral resistance after single-dose nevirapine with tenofovir and emtricitabine to prevent vertical transmission of HIV-1." Aids 23(7): 825-833.

Danel, C., R. Moh, et al. (2009). "Two-Months-off, Four-Months-on Antiretroviral Regimen Increases the Risk of Resistance, Compared with Continuous Therapy: A Randomized Trial Involving West African Adults." Journal of Infectious Diseases 199(1): 66-76.

Danel, C., R. Moh, et al. (2006). "CD4-guided structured antiretroviral treatment interruption strategy in HIV-infected adults in west Africa (Trivacan ANRS 1269 trial): a randomised trial." Lancet 367(9527): 1981-1989.

Darwich, L., A. Esteve, et al. (2008). "Drug-resistance mutations number and K70R or T215Y/F substitutions predict treatment resumption during guided treatment interruptions." Aids Research and Human Retroviruses 24(5): 725-732.

Davey, R. T., N. Bhat, et al. (1999). "HIV-1 and T cell dynamics after interruption of highly active antiretroviral therapy (HAART) in patients with a history of sustained viral suppression." Proceedings of the National Academy of Sciences of the United States of America 96(26): 15109-15114.

Deeks, S. G., M. Smith, et al. (1997). "HIV-1 protease inhibitors - A review for clinicians." Jama-Journal of the American Medical Association 277(2): 145-153.

Dybul, M., E. Nies-Kraske, et al. (2003). "Long-cycle structured intermittent versus continuous highly active antiretroviral therapy for the treatment of chronic infection with human immunodeficiency virus: Effects on drug toxicity and on immunologic and virologic parameters." Journal of Infectious Diseases 188(3): 388-396.

El-Sadr, W. M., J. D. Lundgren, et al. (2006). "CD4+count-guided interruption of antiretroviral treatment." New England Journal of Medicine 355(22): 2283-2296.

Eshleman, S. H., L. A. Guay, et al. (2005). "Distinct patterns of emergence and fading of K103N and Y181C in women with subtype A vs. D after single-dose nevirapine - HIVNET 012." Jaids-Journal of Acquired Immune Deficiency Syndromes 40(1): 24-29.

Eshleman, S. H., D. R. Hoover, et al. (2005). "Nevirapine (NVP) resistance in women with HIV-1 subtype C, compared with subtypes A and D, after the administration of single-dose NVP." Journal of Infectious Diseases 192(1): 30-36.

Farr, S. L., J. A. E. Nelson, et al. (2010). "Addition of 7 Days of Zidovudine Plus Lamivudine to Peripartum Single-Dose Nevirapine Effectively Reduces Nevirapine Resistance Postpartum in HIV-Infected Mothers in Malawi." Jaids-Journal of Acquired Immune Deficiency Syndromes 54(5): 515-523.

Fox, Z., A. Phillips, et al. (2008). "Viral resuppression and detection of drug resistance following interruption of a suppressive non-nucleoside reverse transcriptase inhibitor-based regimen." Aids 22(17): 2279-2289.

Fu, Y. X. (2001). "Estimating mutation rate and generation time from longitudinal samples of DNA sequences." Molecular Biology and Evolution 18(4): 620-626.

Gadhamsetty, S. and N. M. Dixit (2010). "Estimating Frequencies of Minority Nevirapine-Resistant Strains in Chronically HIV-1-Infected Individuals Naive to Nevirapine by Using Stochastic Simulations and a Mathematical Model." Journal of Virology 84(19): 10230-10240.

Gallant, J. E., S. Staszewski, et al. (2004). "Efficacy and safety of tenofovir DF vs stavuldine in combination therapy in antiretroviral-naive patients - A 3-year randomized trial." Jama-Journal of the American Medical Association 292(2): 191-201.

Garcia, F., M. Plana, et al. (2001). "The virological and immunological consequences of structured treatment interruptions in chronic HIV-1 infection." Aids 15(9): F29-F40.

Gardner, E. M., K. H. Hullsiek, et al. (2010). "Antiretroviral medication adherence and class-specific resistance in a large prospective clinical trial." Aids 24(3): 395-403.

Geretti, A. M., Z. V. Fox, et al. (2009). "Low-Frequency K103N Strengthens the Impact of Transmitted Drug Resistance on Virologic Responses to First-Line Efavirenz or Nevirapine-Based Highly Active Antiretroviral Therapy." Jaids-Journal of Acquired Immune Deficiency Syndromes 52(5): 569-573.

Gianella, S. and D. D. Richman (2010). "Minority Variants of Drug-Resistant HIV." Journal of Infectious Diseases 202(5): 657-666.

Goodman, D. D., N. A. Margo, et al. (2009). "Pre-existing low-levels of the K103N HIV-1 RT mutation above a threshold is associated with virological failure in treatment-naive patients undergoing EFV-containing antiretroviral treatment." Antiviral Therapy 14(4): 41.

Haldane, J. B. S. (1927). "A mathematical theory of natural and artificial selection, Part V: Selection and mutation." Proceedings of the Cambridge Philosophical Society 23: 838-844.

Harrigan, P. R., M. Whaley, et al. (1999). "Rate of HIV-1 RNA rebound upon stopping antiretroviral therapy." Aids 13(8): F59-F62.

Hedskog, C., M. Mild, et al. (2010). "Dynamics of HIV-1 Quasispecies during Antiviral Treatment Dissected Using Ultra-Deep Pyrosequencing." Plos One 5(7): 14.

Henry, K., D. Katzenstein, et al. (2006). "A pilot study evaluating time to CD4 T-cell count < 350 cells/mm(3) after treatment interruption following antiretroviral therapy +/- interleukin 2: Results of ACTG A5102." Jaids-Journal of Acquired Immune Deficiency Syndromes 42(2): 140-148.

Hermisson, J. and P. S. Pennings (2005). "Soft sweeps: Molecular population genetics of adaptation from standing genetic variation." Genetics 169(4): 2335-2352.

Hoen, B., I. Fournier, et al. (2005). "Structured treatment interruptions in primary HIV-1 infection - The ANRS 100 PRIMSTOP Trial." Jaids-Journal of Acquired Immune Deficiency Syndromes 40(3): 307-316.

Hudelson, S. E., M. S. McConnell, et al. (2010). "Emergence and persistence of nevirapine resistance in breast milk after single-dose nevirapine administration." Aids 24(4): 557-561.

Jackson, J. B., G. Becker-Pergola, et al. (2000). "Identification of the K103N resistance mutation in Ugandan women receiving nevirapine to prevent HIV-1 vertical transmission." Aids 14(11): F111-F115.

Johnson, J. A. and A. M. Geretti (2010). "Low-frequency HIV-1 drug resistance mutations can be clinically significant but must be interpreted with caution." Journal of Antimicrobial Chemotherapy 65(7): 1322-1326.

Johnson, V. A., F. Brun-VŽzinet, et al. (2010). Update of the Drug Resistance Mutations in HIV-1: December 2010. Top HIV Med. 18 156-163.

Kassaye, S., E. Lee, et al. (2007). "Drug resistance in plasma and breast milk after single-dose nevirapine in subtype C HIV type 1: Population and clonal sequence analysis." Aids Research and Human Retroviruses 23(8): 1055-1061.

Kearney, M., S. Yu, et al. (2009). "HIV-1 plasma virus diversity persists despite suppression with antiretroviral therapy." Antiviral Therapy 14(4): 6.

Kepler, T. B. and A. S. Perelson (1998). "Drug concentration heterogeneity facilitates the evolution of drug resistance." Proceedings of the National Academy of Sciences of the United States of America 95(20): 11514-11519.

Kouyos, R. D., V. von Wyl, et al. (2011). "Ambiguous Nucleotide Calls From Population-based Sequencing of HIV-1 are a Marker for Viral Diversity and the Age of Infection." Clinical Infectious Diseases 52(4): 532-539.

Lallemant, M., N. Ngo-Giang-Huong, et al. (2010). "Efficacy and Safety of 1-Month Postpartum Zidovudine-Didanosine to Prevent HIV-Resistance Mutations after Intrapartum Single-Dose Nevirapine." Clinical Infectious Diseases 50(6): 898-908.

Lee, E. J., R. Kantor, et al. (2005). "Breast-milk shedding of drug-resistant HIV-1 subtype C in women exposed to single-dose nevirapine." Journal of Infectious Diseases 192(7): 1260-1264.

Li, J. Z. and D. R. Kuritzkes (2011). Minority HIV-1 drug resistance mutations and the risk of NNRTI-based antiretroviral treatment failure: a systematic review and pooled analysis.

Lima, V. D., P. R. Harrigan, et al. (2010). "Epidemiology of Antiretroviral Multiclass Resistance." American Journal of Epidemiology 172(4): 460-468.

Lockman, S., R. L. Shapiro, et al. (2007). "Response to antiretroviral therapy after a single, peripartum dose of nevirapine." New England Journal of Medicine 356(2): 135-147.

Loubser, S., P. Balfe, et al. (2006). "Decay of K103N mutants in cellular DNA and plasma RNA after single-dose nevirapine to reduce mother-to-child HIV transmission." Aids 20(7): 995-1002.

Ly, N., V. Phoung, et al. (2007). "Reverse transcriptase mutations in Cambodian CRF01\_AE isolates after antiretroviral prophylaxis against HIV type 1 perinatal transmission." Aids Research and Human Retroviruses 23(12): 1563-1567.

Mansky, L. M. and H. M. Temin (1995). "Lower in vivo mutation rate of human-immunodeficiency virus type 1 than that predicted from the fidelity of purified reverse transcriptase." Journal of Virology 69(8): 5087-5094.

Margot, N. A., B. Lu, et al. (2006). "Resistance development over 144 weeks in treatment-naive patients receiving tenofovir disoproxil fumarate or stavudine with lamivudine and efavirenz in Study 903." Hiv Medicine 7(7): 442-450.

Martinez-Picado, J. and M. A. Martinez (2008). "HIV-1 reverse transcriptase inhibitor resistance mutations and fitness: A view from the clinic and ex vivo." Virus Research 134(1-2): 104-123.

Martinson, N. A., L. Morris, et al. (2009). "Women exposed to single-dose nevirapine in successive pregnancies: effectiveness and nonnucleoside reverse transcriptase inhibitor resistance." Aids 23(7): 809-816.

McIntyre, J. A., M. Hopley, et al. (2009). "Efficacy of Short-Course AZT Plus 3TC to Reduce Nevirapine Resistance in the Prevention of Mother-to-Child HIV Transmission: A Randomized Clinical Trial." Plos Medicine 6(10): 9.

Mocroft, A., S. Vella, et al. (1998). "Changing patterns of mortality across Europe in patients infected with HIV-1." Lancet 352(9142): 1725-1730.

Palmisano, L., M. Giuliano, et al. (2007). "Determinants of virologic and immunologic outcomes in chronically HIV-Infected subjects undergoing repeated treatment interruptions - The Istituto Superiore di Sanita-Pulsed Antiretroviral Therapy (ISS-PART) Study." Jaids-Journal of Acquired Immune Deficiency Syndromes 46(1): 39-47.

Paredes, R., C. M. Lalama, et al. (2010). "Pre-existing Minority Drug-Resistant HIV-1 Variants, Adherence, and Risk of Antiretroviral Treatment Failure." Journal of Infectious Diseases 201(5): 662-671.

Paredes, R., M. Sagar, et al. (2009). "In Vivo Fitness Cost of the M184V Mutation in Multidrug-Resistant Human Immunodeficiency Virus Type 1 in the Absence of Lamivudine." Journal of Virology 83(4): 2038-2043.

Pirillo, M., L. Palmisano, et al. (2010). "Nonnucleoside Reverse Transcriptase Inhibitor Concentrations During Treatment Interruptions and the Emergence of Resistance: A Substudy of the ISS-PART Trial." Aids Research and Human Retroviruses 26(5): 541-545.

Rajesh, L., K. Ramesh, et al. (2010). "Emergence of drug resistant mutations after single dose nevirapine exposure in HIV-1 infected pregnant women in south India." Indian Journal of Medical Research 132(5): 509-512.

Reynolds, S. J., C. Kityo, et al. (2010). "A Randomized, Controlled, Trial of Short Cycle Intermittent Compared to Continuous Antiretroviral Therapy for the Treatment of HIV Infection in Uganda." Plos One 5(4).

Reynolds, S. J., C. Kityo, et al. (2009). "Evolution of drug resistance after virological failure of a first-line highly active antiretroviral therapy regimen in Uganda." Antiviral Therapy 14(2): 293-297.

Ribeiro, R. M., S. Bonhoeffer, et al. (1998). "The frequency of resistant mutant virus before antiviral therapy." Aids 12(5): 461-465.

Rouzine, I. M. and J. M. Coffin (1999). "Linkage disequilibrium test implies a large effective population number for HIV in vivo." Proceedings of the National Academy of Sciences of the United States of America 96(19): 10758-10763.

Ruiz, L., R. Paredes, et al. (2007). "Antiretroviral therapy interruption guided by CD4 cell counts and plasma HIV-1 RNA levels in chronically HIV-1-infected patients." Aids 21(2): 169-178.

Shapiro, R. L., I. Thior, et al. (2006). "Maternal single-dose nevirapine versus placebo as part of an antiretroviral strategy to prevent mother-to-child HIV transmission in Botswana." Aids 20(9): 1281-1288.

Tam, L. W. Y., C. K. S. Chui, et al. (2008). "The Relationship Between Resistance and Adherence in Drug-Naive Individuals Initiating HAART Is Specific to Individual Drug Classes." Jaids-Journal of Acquired Immune Deficiency Syndromes 49(3): 266-271.

Taylor, S., M. Boffito, et al. (2007). "Stopping antiretroviral therapy." Aids 21(13): 1673-1682.
Team, R. D. C. (2005). R: A language and environment for statistical computing. Vienna, Austria, R Foundation for Statistical Computing.

Tomasetti, C. (2011). "A new hypothesis: imatinib affects leukemic stem cells in the same way it affects all other leukemic cells." Blood Cancer Journal 1: e19.

Toni, T. D., B. Masquelier, et al. (2005). "Characterization of nevirapine (NVP) resistance mutations and HIV type 1 subtype in women from Abidjan (Cote d'Ivoire) after NVP single-dose prophylaxis of HIV type 1 mother-to-child transmission." Aids Research and Human Retroviruses 21(12): 1031-1034.

Trkola, A., H. Kuster, et al. (2005). "Delay of HIV-1 rebound after cessation of antiretroviral therapy through passive transfer of human neutralizing antibodies." Nature Medicine 11(6): 615-622.

van Heeswijk, R. P. G., A. Veldkamp, et al. (2001). "Combination of protease inhibitors for the treatment of HIV-1-infected patients: a review of pharmacokinetics and clinical experience." Antiviral Therapy 6(4): 201-229.

van Heeswijk, R. P. G., A. I. Veldkamp, et al. (2000). "The steady-state pharmacokinetics of nevirapine during once daily and twice daily dosing in HIV-1-infected individuals." Aids 14(8): F77-F82.

van Zyl, G. U., M. Claassen, et al. (2008). "Zidovudine with nevirapine for the prevention of HIV mother-to-child transmission reduces nevirapine resistance in mothers from the Western Cape, South Africa." Journal of Medical Virology 80(6): 942-946.

Weis, S. E., P. C. Slocum, et al. (1994). "The effect of directly observed therapy on the rates of drug-resistance and relapse in tuberculosis." New England Journal of Medicine 330(17): 1179-1184.

Yerly, S., C. Fagard, et al. (2003). "Drug resistance mutations during structured treatment interruptions." Antiviral Therapy 8(5): 411-415.

\pagebreak
\section{Supplementary material}
\subsection{Model}
\label{Model}
The model describes population dynamics and population genetics of a panmictic virus population in a single patient. Only virus in infected cells is considered and the dynamics of free virus and uninfected cells are ignored. The total number of cells that can be infected is limited to $K$, the carrying capacity. Because the life cycle of the virus is simplified to just one step in which infected cells ``give birth" to infected cells, all drugs have the same effect: they reduce the number of cells that can be infected by the virus in an infected cell. We will refer to the infected cells by the virus they are infected with. 

When patients fail therapy, this is often due to virus that is resistant against just one drug even if the patient was treated with three drugs (as described in the Introduction). We therefore assume in our model that there are just two types of viruses: wildtype virus that is susceptible to all drugs and resistant virus that is resistant against one of the drugs in the regimen of the patient. We assume that prior to establishment of single drug-resistant virus (hereafter: resistant virus), there is no virus present that is resistant against more than one drug. We only focus on the probability that single drug-resistant virus becomes established.
 
The resistant type has a growth rate larger than $1$ even in the presence of drugs, but in the absence of drugs it is less fit than the wildtype. 
Initially the population consists of only wildtype virus, resistant virus can be created by mutation. Absolute fitnesses  are determined for wildtype and resistant virus, in the absence of drugs (denoted by $u$ for untreated) and during treatment (either ART or NVP).  
These fitnesses are absolute fitnesses, which are equivalent to the reproductive ratio, or the expected number of cells a viral particle would infect in a completely susceptible population of host cells. Without drugs, wildtype virus is fitter than resistant virus, so that the population will always be dominated by wildtype virus. During treatment, resistant virus is fitter than wildtype, so that the population will ultimately become dominated by resistant virus. However, if evolution is mutation-limited, it can take a long time before the equilibrium with resistant virus is reached. One of the successes of combination therapy is to make this waiting time much longer than with monotherapy. 

The relative cost or selective disadvantage is defined as $C_{rel}= (Fwt_{u}-Fm_{u})/(Fwt_{u})$, or, in other words, how much less fit resistant virus is in the absence of drugs, relative to the wildtype. As will be shown later, the fitness of the wildtype with drugs is not important in the current model (although we do have to determine its value for the simulations). 

If the average fitness of the viral population is larger than 1, the viral population grows exponentially until it reaches the carrying capacity ($K$), at which point the population stops growing and stays stable. If the average fitness of the population is below 1, the population shrinks until it reaches $I/(1-F_{wt})$, where $I$ is a fixed number of particles which is added to the population each generation. This reflects the latently infected cells in an HIV infected patient. These latently infected cells have very long half-lifes and function as a reservoir for the HIV population. Because of these cells, the virus population does not die out, even if therapy is very successful and there is (almost) no replication.

\textbf{Simulation details.} Time is counted in generations. Each generation is split in 10 time-steps. At each time step, each viral particle has a probability of $1/10$ to be chosen to reproduce and die. The expected life-time of a viral particle is therefore 1 generation and this is the same for wildtype and resistant virus. 
All particles that are chosen to reproduce at a given time-step die and are replaced by their offspring (i.e. newly infected cells). The number of offspring is equal to $[\text{number of chosen wildtype particles}]*[F_{wt}]+[\text{number of chosen resistant particles}]*[F_{res}]$. If this number is smaller than $20$, stochastic effects should be taken into account, so instead of using the number directly, we take a random number from the Poisson distribution with the expected number as the mean. If adding of the new offspring to the population would lead the population size to be higher than $K$, we add fewer individuals so that the population size will be exactly $K$. The new offspring can be either wildtype or resistant. It is assigned a type at random, with probability $$ \frac{[\text{number of chosen wildtype particles}]*[F_{wt}]} {[\text{number of chosen wildtype particles}]*[F_{wt}]+[\text{number of chosen res particles}]*[F_{res}]}$$ to be wildtype, and resistant otherwise. This is a standard procedure (also used in Hermisson and Pennings 2005)
The choice of 10 time-steps per generation is a compromise between the speed of simulating discrete generations and the reality of continuous replication. 

Mutation occurs only from wildtype to resistant. Back-mutation to wildtype is ignored. The number of mutations in a generation depends on the (per particle and per generation) mutation rate, $\mu$, and the number of newly created infected cells, which in turn depends on fitness and population size. 

\textbf{Analytical calculation of the fixation probability.} 
To calculate the probability that resistance mutations become established when treatment is started (either for the first time or after an interruption), there are two possibilities. When the relative cost of resistance in the absence of drugs, the fitness of the resistant virus with drugs and the viral population size are all known, we can use the theory from Hermisson and Pennings (2005) to calculate the probability that resistance mutations from the standing genetic variation become established. 
Alternatively, if the number of copies of the resistant virus and the fitness of resistant virus are known, it is possible to calculate the probability that at least one of these copies will survive, in which case the population becomes resistant. 
For the latter, we need the probability that a single resistant mutant spreads in the population and ultimately goes to fixation, as opposed to being lost due to genetic drift. 
In the current model, individuals (viral particles in infected cells) will ``produce" a Poisson distributed number of newly infected cells (we call these offspring). The mean number of offspring, $\lambda$, determines the probability that a lineage dies out or survives. Usually, it is assumed that the mean number of offspring of a genotype depends on its fitness relative to the mean fitness of the population. However, when a patient takes drugs, the population size of the virus will be much lower than the carrying capacity, so that there is effectively no competition between resistant and wildtype virus. The mean number of offspring, $\lambda$, therefore does not depend on the relative fitness, but simply equals the absolute fitness of the resistant virus. 
In 1927, Haldane showed that the relationship between the mean number of offspring ($\lambda$) and the fixation probability is 

\begin{equation}
\label{Haldane1}
\lambda =-\frac{log(1-P_{fix})}{P_{fix}}=1+\frac{P_{fix}}{2}+\frac{P_{fix}^2}{3}+\frac{P_{fix}^3}{4}+\frac{P_{fix}^4}{5}+O(P_{fix}^5)
\end{equation}

Traditionally, the fixation probability is calculated using the selection coefficient, $s_b  = \frac{F_{res}-F_{wt}}{F_{wt}}$, instead of the expected number of offspring, $\lambda$. In a population with stable population size and competition between genotypes this is fine because $\lambda=1+s_b$, but in the current model, this does not hold and $s_b$ is irrelevant. 

If $\lambda-1$ is small, a simple approximation for the fixation probability is: 

\begin{equation}
\label{PfixSimple2}
P_{fix}\approx 2(\lambda-1)
\end{equation}

When $\lambda$ is larger, HaldaneÕs equation needs to be solved numerically using more terms of the Taylor expansion.

\subsection{Data from clinical trials}
\textbf{Starting standard therapy.} The probability that resistance mutations from standing genetic variation become established at the start of therapy was estimated from a data set that was published in Margot (2006), see table \ref{TableMargot} for the raw data.
The study reported the number of patients where resistance was detected in the first, second and third year of treatment, for two treatment groups and for three categories of resistance mutations. Because in each case only the first resistance mutation in a given patient was recorded, the data for the three different resistance classes could not be analyzed separately. A saturated model was fitted using R (R Development Core Team, 2005) after which non-significant interactions were removed. It was found that the probability that resistance mutations became established did not differ significantly between the two treatment groups (p=0.26) and also not between year 2 and year 3 (p=0.78). However, there was a clear difference between year 1 and the other two years ($p<10^{-5}$). The average risk per year in year 2 and 3 was 3.7\%, whereas the average risk in year 1 was 9.5\%. The difference between year 1 and the other 2 years was 5.8\%, which could be due to the standing genetic variation before treatment started. 

\textbf{Single dose nevirapine.} We searched the web of science (ISI) database for published studies on single dose nevirapine and resistance. We limited ourselves to studies that reported the fraction of women that had detectable resistance at 4 to 6 weeks postpartum using standard sequencing. An overview of the studies can be found in table \ref{TableNVP}. 

\textbf{Interruption studies.} Seven studies were identified that reported on clinical trials in which patients interrupted treatment multiple times for a fixed length of time (i.e., all patients in the trial were on the same schedule)  and which tested genotypic resistance on plasma samples of all patients in which treatment failed (and ideally of all patients) before and after the trial. Characteristics of the seven studies can be found in table \ref{TableSTI} (supplementary material). The shortest interruption length was two days and the longest 60 days. If possible, we removed patients which had evidence of resistance before the trial started. In all but one of the studies, patients were on ÒtraditionalÓ triple-drug regimens (NNRTI or PI based with two NRTIs). Only in the Staccato study some patients were on a boosted PI regimen (Ananworanich et al, 2003). These patients were taken out of the analysis because resistance seems to evolve much slower in regimens with boosted PI's compared to NNRTI and unboosted PI regimens (Lima, 2008). In two trials, the interruptions had varying lengths (Hoen 2005 and Palmisano 2007), in which case we used the mean length of the interruptions for the plot (these points are in grey in the plot, see figure \ref{STIData}). 

For each study, the fraction of patients ($F$) which did not acquire any resistance mutations during the trial was calculated. The number of treatment interruptions ($T$) which were relevant for the resistance data was calculated. For example, if genotypic resistance tests were done on samples that were obtained during treatment interruption 4, it was assumed that any resistance had occurred due to treatment interruption 1, 2 or 3 so that $T=3$. Furthermore it was assumed that each treatment interruption (TI ) contributed equally to the probability that resistance mutations became established. The probability ($P$) that resistance mutations became established due to a single interruption was calculated as follows: 

\begin{equation}
\label{ProbPerTI}
F=(1-P)^{T} \Rightarrow P=1-F^{1/T}
\end{equation}

Two of the studies also allowed for an estimation of the risk that resistance mutations became established during continuous treatment (Reynolds 2009 and 2010, and Danel 2009). We find a rate of evolution of resistance of 4.3\% and 4.8\% per year (see table \ref{TableSTI}). Using the mean of these two rates, we can estimate the probability that resistance mutations became established during the time on treatment in each of the trials and we corrected the estimated risk of an interruption by subtracting the probability that resistance mutations became established during treatment. 

\subsection{Supplementary tables}

{\tiny
\begin{table}[p]
\caption{Number of patients with at least one resistance mutation detected by the end of the first, second and third year of NNRTI-based antiretroviral therapy. Data from Margot et al (2006).}
\begin{center}
\begin{tabular}{|c|c|c|c|}
\hline
Year&Group&Resistant&Not Resistant \\ \hline
1&TDF&35&264\\ \hline
2&TDF&8&256\\ \hline
3&TDF&9&247\\ \hline
1&d4T&22&279\\ \hline
2&d4T&10&269\\ \hline
3&d4T&11&258\\ \hline
\end{tabular}
\end{center}
\label{TableMargot}
\end{table}%
}

{\tiny
\begin{table}[p]
\caption{Overview of clinical trials with single dose nevirapine which reported the number of patients with nevirapine resistance detected 6 to 8 weeks after treatment.}
\begin{tabular}{|p{3.5cm}|p{1.5cm}|p{3cm}|p{2.5cm}|p{2cm}|p{2.5cm}|}
\hline
Reference and name trial  &Number of patients & Treatment & Treatment code &  Est. prob of establishment of resistance & Remarks \\ \hline
Eshleman 2005,	Malawi			 & 	65	 & 	sdNVP	 & 	sdNVP	 & 	0.69	& 	\\ \hline
Eshleman 2005,	Uganda			 & 	241	 & 	sdNVP	 & 	sdNVP	 & 	0.26	& 	\\ \hline
Farr 2010, 	Malawi			 & 	65	 & 	sdNVP	 & 	sdNVP	 & 	0.34	& 	\\ \hline
Hudelson 2010	,	Uganda			 & 	30	 & 	sdNVP	 & 	sdNVP	 & 	0.43	& 	\\ \hline
Jackson 2000,	Uganda			 & 	15	 & 	sdNVP	 & 	sdNVP	 & 	0.20	& 	\\ \hline
Kassaye 2007,	Zimbabwe			 & 	32	 & 	sdNVP	 & 	sdNVP	 & 	0.35	& 	\\ \hline
Lee 2005,	Zimbabwe			 & 	32	 & 	sdNVP	 & 	sdNVP	 & 	0.34	& 	\\ \hline
Loubser 2006,	South Africa			 & 	44	 & 	sdNVP	 & 	sdNVP	 & 	0.54	& 	Results only for K103N mutation\\ \hline
Ly 2007,	Cambodia 			 & 	35	 & 	sdNVP	 & 	sdNVP	 & 	0.23		& \\ \hline
Martinson 2009,	South Africa,	HIVNET 012	 & 	108	 & 	sdNVP	 & 	sdNVP	 & 	0.38	& \\ \hline	
Martinson 2009,	South Africa,	HIVNET 012	 & 	193	 & 	sdNVP	 & 	sdNVP	 & 	0.46		& \\ \hline
McIntyre 2009,	South Africa			 & 	74	 & 	sdNVP	 & 	sdNVP	 & 	0.59	& 	\\ \hline
Rajesh 2010,	India			 & 	12	 & 	sdNVP	 & 	sdNVP	 & 	0.33	& 	\\ \hline
Toni 2005,	Ivorycoast			 & 	29	 & 	sdNVP	 & 	sdNVP	 & 	0.21	& 	\\ \hline
Farr 2010,	Malawi			 & 	120	 & 	sdNVP + 7 days 3TC/ZDV	 & 	sdNVP/PP	 & 	0.017& \\ \hline		
McIntyre 2009,	South Africa			 & 	164	 & 	sdNVP + 4 days  3TC/ZDV	 & 	sdNVP/PP	 & 	0.097	& \\ \hline	
McIntyre 2009,	South Africa			 & 	168	 & 	sdNVP + 7 days  3TC/ZDV	 & 	sdNVP/PP	 & 	0.073	& 	\\ \hline
Chaix 2007,	Ivorycoast,	ANRS/Ditrame Plus	 & 	63	 & 	ZDV from 36 weeks + sdNVP	 & 	ZDV/sdNVP	 & 	0.33	& 	\\ \hline
Chalermchokcharoenkit 2009,	Thailand			 & 	190	 & 	ZDV 3rd trimester + sdNVP	 & 	ZDV/sdNVP	 & 	0.18	& 	\\ \hline
Chi 2009 and 2007,	Zambia			 & 	166	 & 	ZDV 3rd trimester + sdNVP	 & 	ZDV/sdNVP	 & 	0.25		& \\ \hline
Lallemant 2009,	Thailand,	PHPT2	 & 	222	 & 	ZDV 3rd trimester + sdNVP	 & 	ZDV/sdNVP	 & 	0.064	& 	\\ \hline
Ly 2007,	Cambodia 			 & 	16	 & 	ZDV from 28 weeks + sdNVP	 & 	ZDV/sdNVP	 & 	0.19		& \\ \hline
Shapiro 2006,	Botswana			 & 	155	 & 	ZDV from 34 weeks + sdNVP	 & 	ZDV/sdNVP	 & 	0.45		& \\ \hline
Van Zijl 2008,	South Africa			 & 	76	 & 	ZDV from 34 weeks + sdNVP	 & 	ZDV/sdNVP	 & 	0.17	& 	\\ \hline
Arrive 2010,	Cambodia/Ivorycoast/South Africa,	TEmAA ANRS 12109	 & 	33	 & 	ZDV from enrollment + sdNVP + 1 week TDF/FTC	 & 	ZDV/sdNVP/PP	 & 	0.0	& 	\\ \hline
Dabis 2009,	Cambodia/Ivorycoast/South Africa,	TEmAA ANRS 12109	 & 	37	 & 	ZDV from enrollment + sdNVP + 1 week TDF/FTC	 & 	ZDV/sdNVP/PP	 & 	0.0	& 	\\ \hline
Lallemant 2009,	Thailand,	PHPT4	 & 	222	 & 	ZDV 3rd trim + sdNVP + 1 month ZDV/DDI	 & 	ZDV/sdNVP/PP	 & 	0.0		
& \\ \hline

\end{tabular}
\label{TableNVP}
\end{table}
}

{\tiny
\begin{table}[p]
\caption{Overview of clinical trials with structured treatment interruptions which reported the number of patients with at least one drug resistance mutation detected.}
\begin{tabular}{|p{2.6cm}|p{0.7cm}|p{1cm}|p{1.5cm}|p{1.3cm}|p{1.2cm}|p{1.cm}|p{1.cm}|p{1.3cm}|p{1.3cm}|p{1.3cm}|}
\hline
Reference and name trial & \% on PI &Number of patients & Patients excluded & Patients with genotypic resistance & Fraction not resistant at end of trial & TI's relevant for calculation & Length of TI (days) & Treatment period (days) & Est. prob of evolution of resistance per TI & Corrected prob of evolution of resistance \\ \hline
\emph{Interruption arms} \\ \hline
Reynolds 2009 and 2010, Uganda FOTO arm & 2\% & 57 & 1 (no genotype) & 4 & 52/56 & 72 & 2 & 5 & 0 & 0\\ \hline
Ananworanich 2003,  Staccato WOWO arm & 0\%& 36 & 22 (on boosted PI) 2 (had resistance before trial) & 3 & 9/12 & 17 & 7 & 7 & 0.02 & 0.02\\ \hline
Reynolds 2009 and 2010, Uganda WOWO arm & 6\%& 32 & 3 (left the trial) & 9 & 20/29 & 18 & 7 & 7 & 0.02 & 0.02\\ \hline
Yerly 2003 , SSITT trial & 100\%& 87 & 4 (lost to follow-up) 3 (already resistance before trial) & 11 & 69/80 & 4 & 14 & 56 & 0.04 & 0.03\\ \hline
Hoen 2005, ANRS 100 Primstop trial & 100\%& 26 & 4 (resistance before trial) 1 (resistance in first TI) & 2 & 19/21 & 2 & 14, 28 (mean 21) & 84 & 0.05 & 0.04\\ \hline
Palmisano 2007, ISS\_PART & 25\%& 136 & 16 (estimated number that had resistance before) & 22 & 98/120 & 3 & 28, 28, 56 (mean 37) & 91 & 0.07 & 0.06\\ \hline
Danel 2009, Ivory Coast NCT 00158405 & 10\% & 325 & 10 \tiny (no genotypes) & 76 & 239/315 & 4 & 60 & 91 & 0.07 & 0.06\\ \hline
\emph{Continuous arms}
\\ \hline
Reynolds 2009 and 2010, Uganda continuous arm & 2\%& 51 & 0 & 3 & 48/51 & --  & -- & 504 & -- & 0.043/year\\ \hline
Danel 2009, Ivory Coast NCT 00158405 & 15\% & 110 & 3 \tiny (no genotypes) & 10 & 97/107 & -- & -- & 728 & -- & 0.048/year\\ \hline
\end{tabular}
\label{TableSTI}
\end{table}
}

\end{document}